\documentclass[a4paper,fleqn,usenatbib]{mnras}

\usepackage[T1]{fontenc}
\usepackage{ae,aecompl}

\usepackage{amssymb}	
\usepackage{amsmath}
\usepackage{graphicx}
\usepackage[usenames,dvipsnames]{xcolor}
\usepackage{bm}

\usepackage{xspace}

\usepackage{hyperref}
 \hypersetup{
     colorlinks=true,
     linkcolor=blue,
     filecolor=blue,
     citecolor =olive,      
     urlcolor=cyan,
 }

\usepackage{multirow}

\usepackage{lineno}

\newcommand{\Ddel}{\delta_{\rm D}   }
\newcommand{\MpcOh}{ \,  \mathrm{Mpc}  \, h^{-1} }
\newcommand{\hOMpc}{ \,  \mathrm{Mpc}^{-1}  \, h  }
\newcommand{\nn}{ \nonumber }

\usepackage{scalerel}
\newcommand*{\paral}{\stretchrel*{\parallel}{\perp}}

\newcommand{\beq}{\begin{equation}}
\newcommand{\eeq}{\end{equation}}
\newcommand{\beqa}{\begin{eqnarray}}
\newcommand{\eeqa}{\end{eqnarray}}

\newcommand{\change}[1]{{\textcolor{black}{#1}}\xspace}

\title[   Photometric BAO Reconstruction ]{ Reconstructing the  Baryonic Acoustic Oscillations  \\  in the presence of photo-$z$ uncertainties }

\author[authors]{authors}
\author[K. C. Chan, et al]{
\parbox{\textwidth}{
Kwan Chuen Chan \thanks{E-mail: chankc@mail.sysu.edu.cn (KCC)},
Guoyuan Lu,
Xin Wang}  \\
School of Physics and Astronomy, Sun Yat-Sen University, 2 Daxue Road, Tangjia, Zhuhai, 519082, China \\
CSST Science Center for the Guangdong-Hongkong-Macau Greater Bay Area, SYSU, Zhuhai, 519082, China 
 }

\begin{document}

\label{firstpage}
\pagerange{\pageref{firstpage}--\pageref{lastpage}}
\maketitle



\begin{abstract}

  The reconstruction method has been widely employed to improve the Baryon Acoustic Oscillations (BAO) measurement in spectroscopic survey data analysis. In this study, we explore the reconstruction of the BAO signals in the realm of photometric data.  By adapting the Zel'dovich reconstruction technique, we develop a formalism to reconstruct the transverse BAO in the presence of photo-$z$ uncertainties \change{under the plane-parallel approximation}.  We access the performance of the BAO reconstruction through comoving  $N$-body simulations.  The transverse reconstruction potential can be derived by solving a 2D potential equation, with the surface density and the radial potential contribution acting as the source terms.   The solution is predominantly determined by the surface density.   As is evident in dense samples, such as the matter field, the transverse BAO reconstruction can enhance both the strength of the BAO signals and their cross correlation with the initial conditions. \change{ At $z=0$, the cross correlation is increased by a factor of 1.2 at $ k_\perp = 0.2 \,  \mathrm{Mpc}^{-1}h $ and 1.4 at  $ k_\perp = 0.3 \, \mathrm{Mpc}^{-1}h $, respectively.  }   We contrast the 2D potential results with the 3D Poisson equation solution, wherein we directly solve the potential equation using the position in photo-$z$ space, and find good agreement.  Additionally, we examine the impact of various conditions, such as the smoothing scales and the level of photo-$z$ uncertainties, on the reconstruction results. We envision the straightforward application of this method to survey data.

\end{abstract}


\maketitle

\section{Introduction}

 
Baryonic Acoustic Oscillations (BAO)  \citep{PeeblesYu1970,SunyaevZeldovich1970} is the imprint of the  primordial acoustic features in the distribution of the large-scale structure.  Since the physics for BAO formation is linear and well-understood [e.g.~\cite{BondEfstathiou1984,BondEfstathiou1987,HuSugiyama1996,HuSugiyamaSilk1997,Dodelson_2003}], the sound horizon scale can be computed to high precision, BAO is widely regarded as a standard ruler in cosmology \citep{WeinbergMortonson_etal2013,Aubourg:2014yra}.  Ever since its clear detections in SDSS  \citep{Eisenstein_etal2005}  and 2dFGS \citep{Cole_etal2005}, BAO measurements have been repeated using numerous spectroscopic datasets at different effective redshifts \citep{Gaztanaga:2008xz, Percival_etal2010, Beutler_etal2011, Blake2012_WiggleZ, Anderson_etal2012, Kazin_etal2014, Ross_etal2015, Alam_etal2017, eBOSS:2020yzd, DESI_BAO_2023}.

Because of the large-scale bulk flow, the significance of the BAO feature deteriorates over time \citep{EisensteinSeoWhite2007,CrocceScoccimarro_2008,Matsubara_etal2008,Seo_Einsenstein2007,Ivanov:2018gjr}.  To sharpen the BAO signals, \cite{Eisenstein:2006nk} reconstructed the BAO feature using an algorithm based on the Zel'dovich approximation (ZA) \citep{Zeldovich1970}.  In reconstruction, the objects are moved to their initial position approximately, and this can partly undo the nonlinear damping and reduce the shift in BAO position, resulting in enhancement of the BAO feature. The method has been studied in further details analytically \citep{Padmanabhan:2008dd,Noh_etal2009,Hikage:2019ihj} and numerically \citep{Seo_etal2008,Seo_etal2010, Padmanabhan_etal2012, Burden_etal2014,Burden_etal2015}. Since its first application to survey data in \cite{Padmanabhan_etal2012}, it has been routinely applied to spectroscopic BAO analyses, e.g.~\cite{Anderson_etal2012,Anderson_etal2014,  Kazin_etal2014, Ross_etal2015, Alam_etal2017, eBOSS:2020yzd, DESI_BAO_2023}. Besides the standard Zel'dovich reconstruction method, more advanced method have been developed \citep{Zhu_etal2017, Wang_etal2017, Schmittfull_etal2017, RyuichiroEisenstein_2018,Shi_etal2018,vonHausegger:2021luu,Nikakhtar_etal2022,MaoWang_etal2021,WangZhao_etal2022}. These methods are expected to deliver higher quality reconstruction results especially  when the number density of the sample is high.

On another front, clustering analyses of the imaging datasets are bringing competitive results. Although the redshifts (photo-$z$'s) are less accurate, imaging surveys are capable of measuring large volume of data with deep magnitude efficiently. This is especially advantageous for BAO measurement because it is a large-scale feature and hence requires big dataset to get a good statistic. Photo-$z$'s are typically so large that the radial BAO cannot be detected on photometric data, but the transverse BAO can  still be  reliably measured \citep{SeoEisenstein_2003,BlakeBridle_2005,Amendola_etal2005,Benitez:2008fs,Zhan:2008jh, Chaves-Montero:2016nmw, Ross:2017emc, Chan:2018gtc,Chan_xiptheory2022,Ishikawa_etal2023}.  Up till now,  photometric BAO measurements have been presented in various photometric data analyses \citep{Padmanabhan_etal2007,EstradaSefusattiFrieman2009, Hutsi2010, Seo_etal2012, Carnero_etal2012, deSimoni_etal2013, Abbott:2017wcz, DES:2021esc, Chan_xip2022}. In particular, the most precise photometric BAO measurement so far is obtained by the DES Y3 \citep{DES:2021esc, Chan_xip2022}.  There are a number of ongoing or upcoming large-scale structure surveys expected to deliver enormous amount of photometric data including the DES Y6, Vera C. Rubin Observatory (or LSST) \citep{Ivezic_etal2019},  the Euclid Space Mission \citep{Laureijs_etal2011},  Nancy Grace Roman Space Telescope (or WFIRST) \citep{Akeson_WFIRST2019}, and the Chinese Space Station Telescope (CSST) \citep{Zhan_2011, Gong_etal2019}. Hence, we anticipate that the photometric BAO measurements will play a bigger role in future.


So far all those photometric BAO measurements have been performed without reconstruction.  With the successful measurements of the transverse BAO from photometric data, it is timely to consider reconstruction of the photometric BAO.
The goal of this paper is to explore the reconstruction of the BAO signals in the large-scale structure photometric data.  The rest of the paper is organized as follows.  We first present a formalism suitable for the transverse BAO reconstruction in the presence of photo-$z$ uncertainties in Sec.~\ref{sec:formalism}.  Using  $N$-body simulation data, we check the performance of the reconstruction method and test how it works under different conditions in Sec.~\ref{sec:NumericalAnalysis}.  Sec.~\ref{sec:conclusions} is devoted to the conclusions.  In Appendix~\ref{appendix:transverse_Pk_xi}, we show how the transverse power spectrum and the transverse correlation function are related to the underlying 3D power spectrum.

\section{Formalism }
\label{sec:formalism}

For spectroscopic data, the BAO reconstruction method is performed using 3D position. For the photometric data, due to the photo-$z$ uncertainties, the radial direction information is substantially smeared out, while the transverse one remains largely intact. Thus we anticipate that BAO reconstruction is effective only in the transverse direction in the photometric case.  In this section, we first review the standard BAO reconstruction and then consider its adaptation for photometric data.

\subsection{ Review of the standard BAO reconstruction }

Here we  review the ZA BAO reconstruction method \citep{Eisenstein:2006nk}.  In ZA reconstruction, the displacement potential $\phi$ is estimated from the Eulerian density field via the Poisson equation
\beq
\label{eq:Poisson_3D_straight}
\nabla^2 \phi = \delta. 
\eeq
To the ZA order, the displacement potential coincides with the gravitational potential. In theory, the density should be the linear one.  In practice the nonlinear density is used as a proxy for the linear one, with the small scale power suppressed by a smoothing window such as a Gaussian.   The Zel'dovich displacement field $\bm{\Psi} $ and  the associated reconstruction displacement $\bm{s}$ are derived from the \change{smoothed}  potential as
\begin{align}
\label{eq:ZA_Psi}
\bm{s}    & = - \bm{\Psi} \change{ \simeq }     \nabla \phi . 
\end{align}

In the standard BAO reconstruction, both the displace and shift catalogs are required to generate the reconstructed field.  In the displace catalog, the objects (matter particles or halos) are displaced by $  \bm{s} $ to their initial position approximately. This not only reduces the impact of nonlinear evolution, but also suppresses the power on large scale.  The shift catalog is constructed by shifting particles at the grid position by $ \bm{s} $. It is designed to restore the large-scale power of the displace catalog.  The reconstructed overdensity  is given by
\beq
\delta_{\rm rec} = \delta_{\rm displace } - \delta_{ \rm shift},
\eeq
where  $\delta_{\rm displace }$ and $ \delta_{ \rm shift} $ are the density contrast of the displace and shift catalogs, respectively.  Indeed, as demonstrated by \cite{Padmanabhan:2008dd}, $ \delta_{\rm rec} $ results in the linear density to first order, but  it deviates from linear theory at second order.

\subsection{ 3D Poisson equation inversion }
\label{sec:Direct3D_inversion}
After reviewing the standard  BAO reconstruction procedures for spectroscopic data, we now move to discuss the case for photometric data.  We now show that we can simply treat the density in photo-$z$ space as a  3D field and invert the 3D Poisson equation directly to get the displacement potential.

Here we present the field equation in photo-$z$ space. The impact of the photo-$z$ is to probabilistically map the position in true-$z$ space to photo-$z$ space. Statistically, the effect is to smear the field with some distribution function.  To model the effect of the photo-$z$ uncertainty, we project the fields along the radial direction by convolving them with a window function $W$ \citep{CrocceCabreGazta_2011,Chan:2018gtc}. With the window function convolution,  from Eq.~\eqref{eq:Poisson_3D_straight} we get 
\begin{align}
  \label{eq:3D_Poisson_photoz}
  \nabla^2  \phi_{\rm p}( \bm{x}_\perp, x_{\paral}^{\rm p})  =  \delta_{\rm p}( \bm{x}_\perp, x_{\paral}^{\rm p})  ,
\end{align}
where
\begin{align}
\phi_{\rm p}( \bm{x}_\perp,x_{\paral}^{\rm p} ) &=   \int d x_{\paral}   W(  x_{\paral}^{\rm p} - x_{\paral}  ) \phi( \bm{x}_\perp , x_{\paral} ) , \\
\delta_{\rm p}( \bm{x}_\perp, x_{\paral}^{\rm p}  ) & = \int d x_{\paral}  W( x_{\paral}^{\rm p} - x_{\paral}   ) \delta ( \bm{x}_\perp , x_{\paral} ) ,
\end{align}
where $ \bm{x}_\perp $ is the transverse coordinates, while $x_{\paral} $ and  $x_{\paral}^{\rm p} $  denote the radial coordinates \footnote{Although we take the plane parallel limit in this work, we still call it the radial direction. } in true position space and photo-$z$ space, respectively.   In this work, we assume the distant observer approximation, and leave the study with the curved sky effect for future work.   Note that in Eq.~\eqref{eq:3D_Poisson_photoz}, the radial Laplacian acts on $ x_{\paral}^{\rm p}$ because  we can express  the radial Laplacian contribution as
\begin{align}
  & \int d x_{\paral} W( x_{\paral}^{\rm p} - x_{\paral} ) \frac{\partial^2}{\partial x_{\paral}^2} \phi( \bm{x}_\perp , x_{\paral} )  \nn  \\
  & =  \frac{\partial^2}{\partial x_{\paral}^{\rm p 2} }  \int d x_{\paral}  \phi( \bm{x}_\perp , x_{\paral} )    W( x_{\paral}^{\rm p} - x_{\paral} ) ,
\end{align}
by  applying the integration by parts and then exploiting the symmetry between  $x_{\paral}^{\rm p}$ and $x_{\paral} $.  Eq.~\eqref{eq:3D_Poisson_photoz} shows that the potential in photo-$z$ space is still governed by the Poisson equation, except with the quantities replaced by the ones in photo-$z$ space. This indeed demonstrates that we can straightforwardly generalize  the spec-$z$ 3D approach to the photometric data case.

While Eq.~\ref{eq:3D_Poisson_photoz} holds for any reasonable photo-$z$ window function, we shall take it to be a Gaussian in the numerical calculations.  We comment that photo-$z$ is a stochastic process, and hence on small scales there are random fluctuations on top of the deterministic convolution given above. The convolution method is a good approximation only when the particle number density is large.  The stochastic fluctuations can be suppressed by applying a smoothing window, although aggressive smoothing will also damp the small-scale signal. We shall study the impacts of the stochasticity later on.

After getting the potential  $\phi_{\rm p} $ by  inverting Eq.~\eqref{eq:3D_Poisson_photoz}, we can use it as the displacement potential (see Sec.~\ref{sec:TransverseDisplacement} for justifications). To reduce the impact of the radial direction, which is substantially contaminated by photo-$z$, we then project the potential to the transverse direction.



In Sec.~\ref{sec:various_solution_check}  we  check this method against simulation, and find that this method indeed works.  Nonetheless, it is not entirely clear how the photo-$z$ impacts the reconstruction, in particular how important the radial contribution is.  To gain more insights on this issue, in next subsection, we pay particular attention to isolate the information in the transverse direction and derive a potential equation that is particularly appropriate for transverse BAO reconstruction.

\subsection{ Transverse potential equation } 
\label{sec:transverse_potential_eq}
 As the photo-$z$ uncertainties primarily affect the radial direction, we now explicitly split Eq.~\eqref{eq:3D_Poisson_photoz} into radial and transverse direction as 
 \beq
 \label{eq:Poisson_split_photoz}
  \left(\nabla_{\perp}^2+\frac{\partial^2}{\partial x_{\paral}^2}\right) \phi_{\rm p} ( \bm{x}_\perp , x_{\paral} )   =    \delta_{\rm p} ( \bm{x}_\perp , x_{\paral} )  ,
  \eeq
  where $\nabla_{\perp}^2 $ denotes the Laplacian operator in the transverse direction.

As in Eq.~\eqref{eq:3D_Poisson_photoz}, we can rewrite Eq.~\eqref{eq:Poisson_split_photoz}  as 
  \begin{align}
  \label{eq:eff_2D_Poisson_bare}
  \nabla_{\perp}^2  \phi_{\rm p}( \bm{x}_\perp, x_{\paral}^{\rm p})   +  \frac{\partial^2}{\partial x_{\paral}^{\rm p 2} } \phi_{\rm p}( \bm{x}_\perp, x_{\paral}^{\rm p})    =  \delta_{\rm p}( \bm{x}_\perp, x_{\paral}^{\rm p})  .
\end{align}
  If we move the radial contribution term to the RHS, it effectively acts as an additional source to the Poisson equation.  Even if we limit ourselves to the density on a 2D surface,  we still get this remnant effect from the 3D motion. However, we shall see that this term is negligible.


  In practice, in clustering analyses of the photometric data, we often divide them into tomographic bins of thickness comparable to the photo-$z$ uncertainty.  Here we consider a tomographic bin of width $L$ and average $\phi_{\rm p} $ and $\delta_{\rm p}$ over $L$ to get
  \begin{align}
\label{eq:eff_2D_Poisson_slabave}    
    \nabla_{\perp}^2   \langle \phi_{\rm p} \rangle  ( \bm{x}_\perp )   =   \langle  \delta_{\rm p}  \rangle ( \bm{x}_\perp ) -  \Big\langle  \frac{ \partial^2 }{ \partial x_{\paral}^{\rm p 2} }   \phi_{\rm p}  ( \bm{x}_\perp, x_{\paral}^{\rm p} ) \Big\rangle  ,
\end{align}
  where  the angular bracket represents averaging of the quantity over the slab. Explicitly, the last term denotes 
  \beq
 \Big\langle  \frac{ \partial^2}{\partial x_{\paral}^{\rm p 2} }   \phi_{\rm p}  ( \bm{x}_\perp, x_{\paral}^{\rm p} ) \Big\rangle =         \frac{1}{L}  \int_L dx_{\paral}^{\rm p}   \frac{ \partial^2 }{ \partial  x_{\paral}^{\rm p 2} }   \phi_{\rm p}  ( \bm{x}_\perp, x_{\paral}^{\rm p} )   .
 \eeq
It is tempting to simplify it further to 
\beq
\label{eq:phipp_smoothintg}
 \Big\langle  \frac{\partial^2}{\partial x_{\paral}^{\rm p 2} }   \phi_{\rm p}  ( \bm{x}_\perp, x_{\paral}^{\rm p} ) \Big\rangle  =    \frac{1}{L} \frac{\partial}{\partial x_{\paral}^{\rm p } }   \phi_{\rm p}  ( \bm{x}_\perp, x_{\paral}^{\rm p} ) \bigg|_{  x_{\paral,0}^{\rm p}  }^{  x_{\paral,1}^{\rm p}  } , 
 \eeq
where  $ x_{\paral,0}^{\rm p}$  and  $x_{\paral,1}^{\rm p}$ denote boundaries of the tomographic bin. 
This implicitly assumes the smoothness of $ \frac{\partial^2}{\partial x_{\paral}^{\rm p 2} }   \phi_{\rm p}  $ so that interior contribution cancels out, leaving only the boundary terms.  However, the stochasticity violates the smoothness assumption, invalidating  Eq.~\eqref{eq:phipp_smoothintg}. Thus we average over the second derivative term directly without resorting to Eq.~\eqref{eq:phipp_smoothintg}.

We solve Eq.~\eqref{eq:eff_2D_Poisson_slabave} by regarding the RHS terms as the source terms. Because photo-$z$ smearing causes the field to be coherent, the magnitude of the second derivative term along the radial direction is significantly suppressed relative to the transverse ones. It is further reduced by averaging over the slab width.  Consequently, $  \Big\langle  \frac{\partial^2}{\partial x_{\paral}^{\rm p 2} }   \phi_{\rm p} \Big\rangle $ is subdominant compared to  $\langle \delta_{\rm p}  \rangle $. Our strategy is to take the $ \phi_{\rm p} $ solution from Eq.~\eqref{eq:3D_Poisson_photoz} and plug it into Eq.~\eqref{eq:eff_2D_Poisson_slabave} as the source term.

\subsection{ Biased tracers }
\label{sec:transverse_eq_biased}

In the previous subsection, we assume that the tracer is unbiased. For biased tracer, Eq.~\eqref{eq:Poisson_split_photoz} is generalized to 
\beq
\label{eq:eff_2D_Poisson_bare_biased}
  \left(\nabla_{\perp}^2+\frac{\partial^2}{\partial x_{\paral}^2}\right) \phi ( \bm{x}_\perp , x_{\paral} )   =   \frac{ \delta_{\rm g }  ( \bm{x}_\perp , x_{\paral} )  }{  b } , 
  \eeq
  where $\delta_{\rm g } $ is the overdensity of the biased tracer, such as galaxies, and  $b$ is their linear bias.
  
By going through the same arguments as before, we arrive at 
\begin{align}
  \nabla_{\perp}^2   \langle{\phi}_{\rm p} \rangle ( \bm{x}_\perp )  & =  \frac{  \langle{ \delta}_{\rm g p} \rangle  ( \bm{x}_\perp ) }{b}  -
 \Big\langle  \frac{\partial^2}{\partial x_{\paral}^{\rm p 2} }   \phi_{\rm p}  ( \bm{x}_\perp, x_{\paral}^{\rm p} ) \Big\rangle 
   \label{eq:eff_2D_Poisson_bin_biased}
\end{align}

\subsection{ RSD }

\label{sec:transverse_eq_RSD}

As photo-$z$ uncertainties, redshift space distortion (RSD) also affects the radial direction. However, its effect is typically much weaker than the photo-$z$ smearing.   When RSD is included, Eq.~\eqref{eq:eff_2D_Poisson_bare_biased} is further generalized to \citep{NusserDavis1994,Padmanabhan_etal2012,Seo_etal2016}          
\beq
\label{eq:eff_2D_Poisson_bare_biased_RSD}
  \left(\nabla_{\perp}^2 + \frac{\partial^2}{\partial x_{\paral}^2}\right) \phi ( \bm{x}_\perp , x_{\paral} )    +  \beta  \frac{\partial^2}{\partial x_{\paral}^2} \phi ( \bm{x}_\perp , x_{\paral} )     =   \frac{ \delta_{\rm g }  ( \bm{x}_\perp , x_{\paral} )  }{  b } , 
  \eeq
 where $\beta = f /b $, with $f$ being the linear growth rate defined as  $ d \ln D/ d \ln a  $ ($D$ is the linear growth factor).

 We immediately see that the effect of the RSD can be easily accounted for by applying the replacement,  $ 1 \rightarrow (1+\beta )$,  to the factor in front of  $  \frac{\partial^2}{\partial x_{\paral}^2} \phi ( \bm{x}_\perp , x_{\paral} ) $.  Thus we arrive at
\begin{align}
  \nabla_{\perp}^2   \langle \phi_{\rm p} \rangle ( \bm{x}_\perp )  & =  \frac{  \langle \delta_{\rm g p} \rangle  ( \bm{x}_\perp ) }{b}  - ( 1 + \beta )  \Big\langle  \frac{\partial^2}{\partial x_{\paral}^{\rm p 2} }   \phi_{\rm p}  ( \bm{x}_\perp, x_{\paral}^{\rm p} ) \Big\rangle .
   \label{eq:eff_2D_Poisson_bin_biased}
\end{align}

\subsection{ Transverse displacement}
\label{sec:TransverseDisplacement}

Using  the displacement potential \change{ $\phi$   estimated from the late time smoothed density field}, the 3D Zel'dovich displacement is given by
\begin{align}
 \bm{\Psi}  = -  \nabla \phi .  
\end{align}

Limiting to the displacement in transverse direction and  moving to the photo-$z$ space, we have
\begin{align}
  \bm{\Psi}_\perp ( \bm{x}_\perp, x_{\paral}^{\rm p} ) & = \int d x_{\paral} W( x_{\paral}^{\rm p } - x_{\paral} )  \bm{\Psi}_\perp( \bm{x}_\perp, x_{\paral} )  \nn \\
  & =   -   \bm{\nabla}_\perp    \phi_{\rm p}(  \bm{x}_\perp, x_{\paral}^{\rm p} ) .
 \end{align}
 Further averaging over the photo-$z$ tomographic bin width yields
 \beq
 \langle \bm{\Psi}_\perp \rangle  ( \bm{x}_\perp ) =   -  \bm{\nabla}_\perp    \langle \phi_{\rm p} \rangle ( \bm{x}_\perp )  . 
 \eeq
As expected, the gradient of the transverse potential   $ \langle \phi_{\rm p}\rangle $ gives the Zel'dovich displacement, and hence the corresponding transverse reconstruction field  $ \bm{s}_\perp =  - \langle \bm{\Psi}_\perp \rangle $.





\begin{figure*} 
\centering
\includegraphics[width=0.96\linewidth]{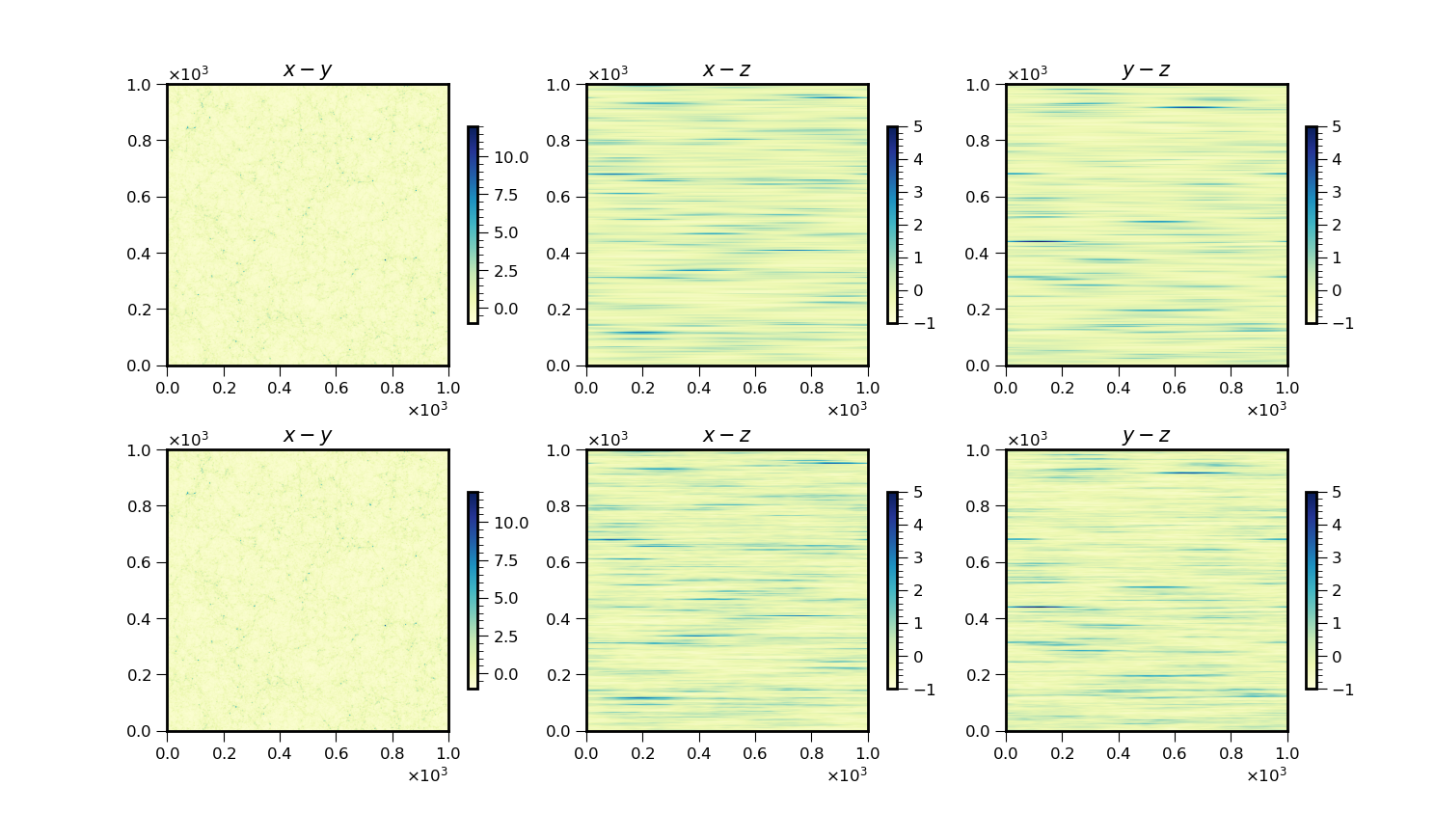}
\caption{ Section of the matter overdensity field in the $x$-$y$, $x$-$z$, and $y$-$z$ plane. The thickness of the section is 1.95 $\MpcOh$. The upper panels show the density obtained deterministically by convolving the true density with a window function, while the results in the lower panels are the density in photo-$z$ space from simulation. The latter results have been smoothed by a Gaussian window with standard deviation of  $ 15 \MpcOh $. The density convolution results and the simulation measurements agree  with each other well. \change{ We have set the same range of scales to ensure consistent contrast between the convolution and simulation measurement results. }  }
\label{fig:densityDM_convol_simG15}
\end{figure*}

\begin{figure*}
\centering
\includegraphics[width=0.96\linewidth]{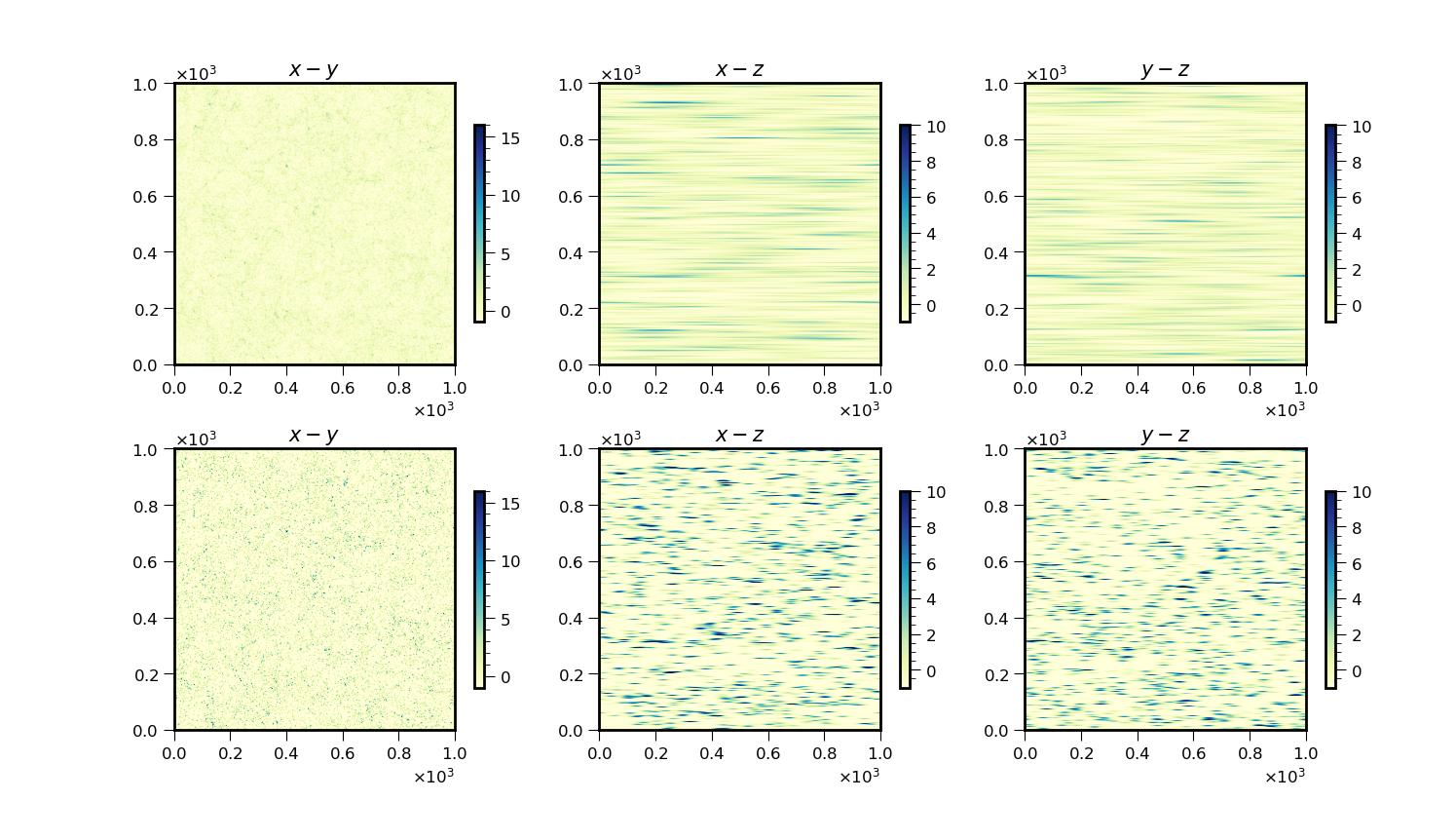}
\caption{ Same as Fig.~\ref{fig:densityDM_convol_simG15}, except for the halo overdensity field. Compared to matter, the impact of stochasticity is much larger. There are significant visual differences between the convolution results and the direct measurements.  } 
\label{fig:densityHalo20_5000_convol_simG15}
\end{figure*}

\section{ Numerical analysis }
\label{sec:NumericalAnalysis}
In this section, we check the performance of the transverse BAO reconstruction using comoving $N$-body simulations. We show the results for both dark matter and halos. Among other things, we will demonstrate the efficacy of the BAO reconstruction method and check how it depends on different conditions. 


We first outline the basic information of the simulations suits, which  were first created  to measure the BAO via the volume statistics \citep{ChanHamaus_2021}. The cosmology of the simulations is a flat $\Lambda$CDM model with parameters $\Omega_{\rm m}=0.3  $,  $\Omega_{\Lambda}=0.7  $, $h =0.7$, $n_{\rm s} = 0.967$, and  $\sigma_8=0.85 $. The simulation contains  $ 512^3 $ particles in a cubic box of comoving side length  $ 1000 \MpcOh $.  The initial conditions are Gaussian and they are generated by {\tt CLASS} \citep{CLASS_code} at redshift $z=49$.  The initial particle displacements are implemented using {\tt 2LPTic}~\citep{Crocce:2006ve} and the simulation is evolved with the treePM code {\tt Gadget2} \citep{Springel:2005mi}.  We ran two sets of simulation: one with normal BAO signals and another with Eisenstein-Hu initial conditions without the BAO wiggles \citep{EisensteinHu_1998} (no-wiggle).  Altogether, there are 20 realizations for each set.  Unless otherwise stated, the error bars are the standard error of the mean of the measurements among the realizations.  We shall illustrate our results at $z=1$ and $0$.  We use halos generated by the halo finder {\tt AHF}~\citep{Knollmann:2009pb}. They are defined via a spherical overdensity threshold of 200 times the background density.  In order to maximize the number density, we consider a halo sample with 20 or more particles.  The bias and number density of this sample are 2.63 and $1.61 \times 10^{-4} (\MpcOh)^{-3} $  at $z=1$ and  1.37 and $3.82 \times 10^{-4} (\MpcOh)^{-3} $   at $z=0$.

To model the impact of photo-$z$, we add photo-$z$ uncertainties to the comoving simulation catalog. Independent photo-$z$ uncertainties are assigned randomly to the $z$ coordinate of the objects.   We assume Gaussian photo-$z$ of standard deviation  $ (1+z) \sigma_z $ with $ \sigma_z = 0.03 $, which is the typical photo-$z$ uncertainty attained by wide band imaging surveys, e.g.~\cite{CarneroRosell_etal2022}.  These translate to uncertainties in the comoving scale of 102.2 $\MpcOh$ at $z=1 $ and  89.9 $\MpcOh$  at $z=0 $, respectively.

We will also consider the effect of RSD. To do so, we first add RSD displacement $v_z /(aH)$, where $v_z$ is $z$-component of the peculiar velocity, to the $z$-coordinates, and then further add the photo-$z$ displacement. The default results are in redshift space.

In our fiducial set-up, the $z$ Cartesian direction is divided into 5 slabs  of equal width of 200 $ \MpcOh$, so that each slab is of dimension $ 1000 \times  1000 \times 200  \, (\MpcOh)^3 $.    We  use the transverse power spectrum computed on the slabs to showcase the BAO reconstruction performance.   We take the slab results to be independent measurements, and average over them as well.   

\subsection{ Stochasticity of photo-$z$ }

In our modeling, e.g.~Eq.~\eqref{eq:3D_Poisson_photoz}, we account for the effect of the photo-$z$ uncertainty by convolving the field with a window function.  \change{In this way, given a density field, the corresponding field in photo-$z$ space is deterministically obtained from the convolution operation, while the full photo-$z$  effects are stochastic. }  This deterministic approach  only models the mean effect of the photo-$z$ uncertainty, and it is valid for sizable number of particles so that the magnitude of the stochasticity is suppressed.  Here we check the impact of photo-$z$ stochasticity by comparing the deterministic results against direct measurements from simulation.


In Figs.~\ref{fig:densityDM_convol_simG15} and \ref{fig:densityHalo20_5000_convol_simG15}, we show respectively the sections of the matter and halo overdensity field at $z=0$.  The sections are of width of 1.95 $\MpcOh$. \change{ Note that we have smoothed the fields with a Gaussian window of width 15 $\MpcOh $, and so the results are similar if a large slab width, e.g.~15 $\MpcOh $ is used. }   Because the photo-$z$ uncertainty is applied in the $z$-direction, the structures in $x$-$z$  and $y$-$z$ planes are smeared in $z$ direction, while the ones in $x$-$y$ plane remain largely intact.  

In each plot, we compare the results obtained deterministically by convolving the  underlying redshift space overdensity field with a window function and the direct measurement from simulations in photo-$z$ space. The window function is a Gaussian window of width 89.9 $\MpcOh$, which corresponds to the photo-$z$ uncertainty we inject. For the direct measurement from simulations, we smooth the results with a Guassian window with width 15 $\MpcOh $, which is the fiducial smoothing scale we adopt in reconstruction.

\change{ It is helpful to highlight the differences between the two window functions used here.
We apply a convolution window in the radial direction to model the effect of the photo-$z$ uncertainty.  In first approximation, it is often taken to be Gaussian in form. This smoothing scale is quite large, around $\sim 100 \MpcOh $ for typical broadband imaging surveys.  On the other hand,  when we compute the displacement potential, a smoothing window is applied to suppress the small-scale power so as to approximate the linear density field. This window is isotropic, and it acts on three (two) dimensions for a 3D (2D) density field. The Gaussian window form is often adopted in this case as well, and the smoothing scale usually lies in the range of $ [10, \,  20] \, \MpcOh$. }

For matter field, as the number density is very high, this smoothing is well sufficient as we find that the results obtained with these two means are in good agreement with each other. However, it proves inadequate for the halo sample. Visually, there are substantial differences between these two results on small scales.  \change{ We find that the convolution results are much more smooth, while the simulation measurements are spiky in this case. }  When more aggressive smoothing is applied, e.g.~30 $\MpcOh $, the agreement can get much better.   \change{ However, we point out that in reconstruction, what we really need is the derivative of the potential,  which goes like  $ \delta /k_\perp$, therefore a greater contribution from low $k_\perp$ (large scale). Thus the impact of stochasticity that dominates on small scales is expected to be milder in the displacement estimation than that shown in Figs.~\ref{fig:densityDM_convol_simG15} and \ref{fig:densityHalo20_5000_convol_simG15}. }    We will discuss more on the impact of the stochasticity and the smoothing scale on BAO reconstruction later on.


\begin{figure*}
\centering
\includegraphics[width=0.98\linewidth]{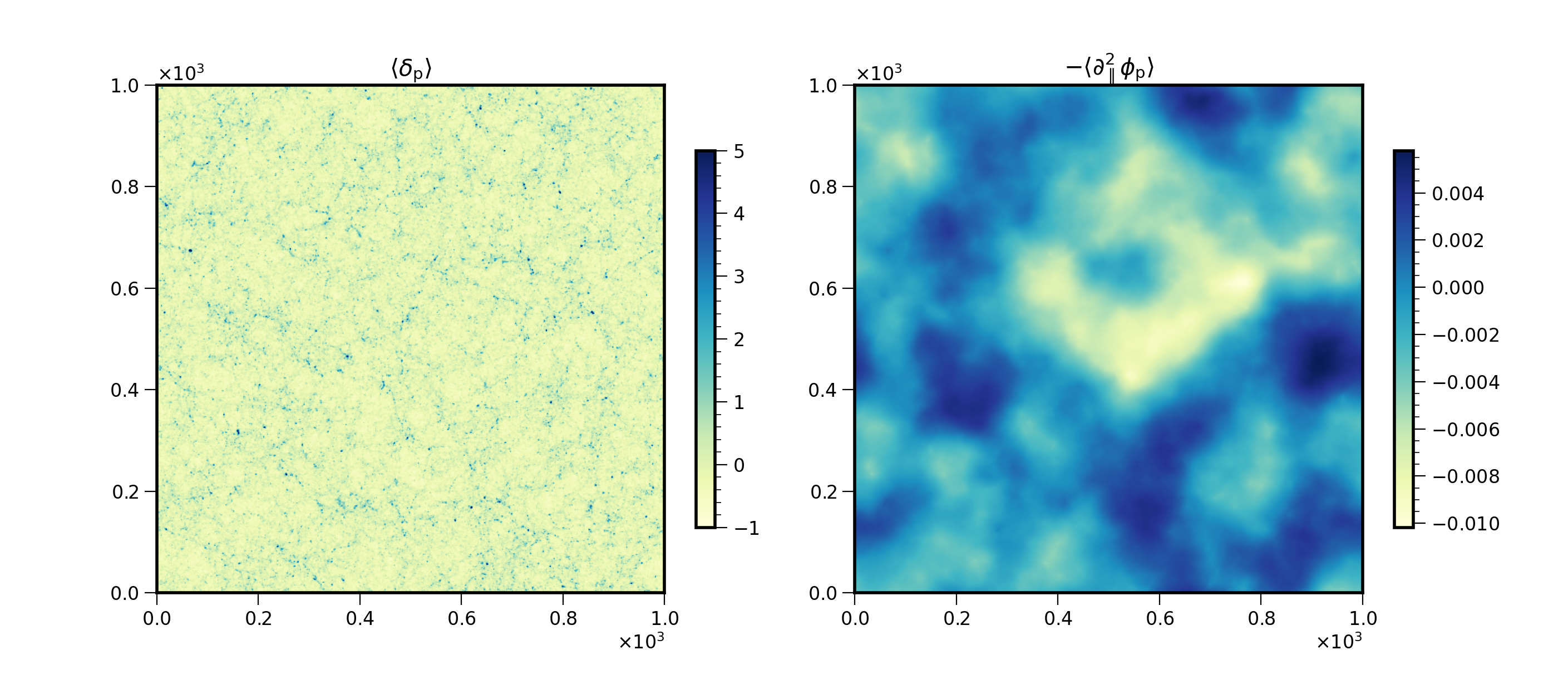}
\caption{ Sources to the transverse potential equation for the matter field at $z=0$. The magnitude of the surface density (left panel) and the radial potential (right panel) term are compared. The data results from a projection of a slab of width of 200 $\MpcOh $. \change{For the overdensity, we have limited the maximum range (from 13 to 5) to increase the contrast. }   The scales indicate that the surface density is completely dominant over the radial potential.   However, because the long wavelength modes play a much bigger role in the solution of the Poisson equation, the radial potential term is not as negligible as it may seem. }
\label{fig:sources_delta2_Dpartialphi_DM}
\end{figure*}

\begin{figure*}
\centering
\includegraphics[width=0.98\linewidth]{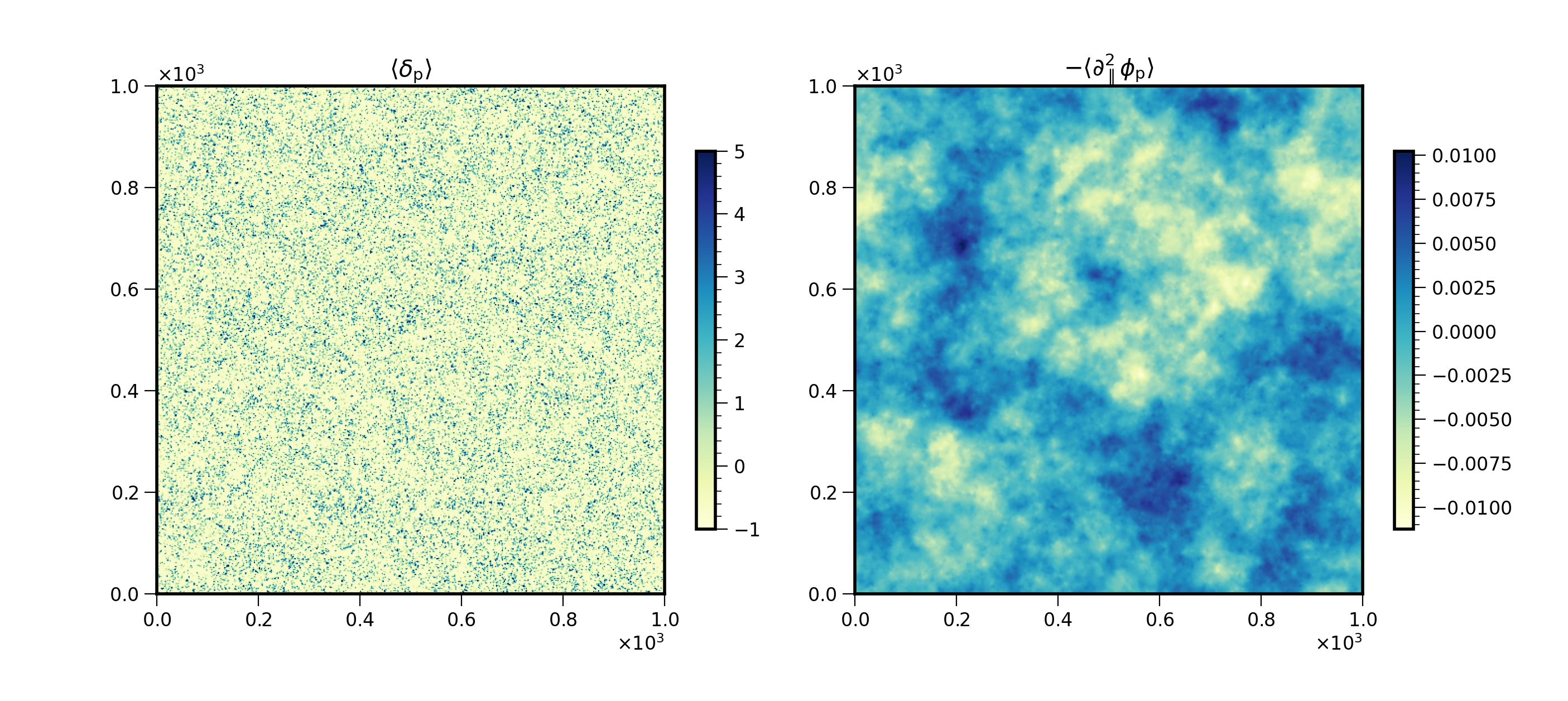}
\caption{ Same as Fig.~\ref{fig:sources_delta2_Dpartialphi_DM} except for the halo field. The stochasticity is much stronger in this case.  }
\label{fig:sources_delta2_Dpartialphi_Halo20_5000}
\end{figure*}

\subsection{ Magnitude of the sources }
\label{sec:sources_magnitude}

It is instructive to check the magnitude of the source terms of the transverse potential equation: $\langle \delta_{\rm p} \rangle  $ and the radial potential term $- \langle  \partial^2_{ \parallel} \phi_{\rm p} \rangle  $.  In Figs.~\ref{fig:sources_delta2_Dpartialphi_DM} and \ref{fig:sources_delta2_Dpartialphi_Halo20_5000}, we show them for the matter and the halo field, respectively.  The data is a slab of thickness of $200 \MpcOh $ taken from simulation at $z=0$.  Both the density field and the potential field have been smoothed by a Gaussian window of $15 \MpcOh$.

From the range of the scale bars, we see that the fluctuations of the density are larger than the radial potential contribution by three orders of magnitude, and this suggests that the reconstruction is dominated by the surface density contribution. We will see that this guess is indeed correct.   Nonetheless, while the visual inspection of the scale is useful to gain intuitions on these terms, it is important to remember that only the long wavelength modes of the sources contribute significantly to the potential thanks to the magnification on the  long wavelength modes by the inverse Laplacian operator. Thus the radial potential term is not as negligible as it may seem.

Although for the halo density, we have  divided by the bias factor, the range of values in the case of halo is still larger than that of the matter.  The deviation from linear biasing can be attributed to the  nonlinearity, non-locality, and stochasticity of the biasing [see e.g.~\cite{Desjacques:2016bnm}]. In particular, the stochasticity in the halo field is  much stronger for halo field than for the matter field.


As we mentioned previously in Sec.~\ref{sec:transverse_potential_eq}, the smallness of the radial term is due to the photo-$z$ smoothing, which smears the field and causes it to be smooth in the $z$ direction, and consequently, the second derivative of the potential is suppressed. This also causes the radial term to be further reduced by the averaging operation relative to the density term.


\subsection{ Reconstruction tests }
\label{sec:ReconstructionTests}

In this subsection, we test the performance of the BAO reconstruction.  To reconstruct the BAO, we follow the ZA BAO reconstruction procedure \citep{Eisenstein:2006nk} to derive the reconstructed catalog from the displace and the shift ones. In the displace catalog, the objects are displaced by $ \bm{s}_\perp $ to their initial position approximately.  This operation decreases  power in the displace catalog including that on large scale.  The shift catalog is built by shifting particles on a grid by  $  \bm{s}_\perp $. We consider a 2D grid of size  $512 \times 512 $ for the shift catalog.  The reconstructed 2D density field is given by
\beq
\delta_{2  \rm rec }  =  \delta_{2 \rm displace} - \delta_{2 \rm shift} , 
\eeq
where  $ \delta_{2 \rm displace} $ and $ \delta_{2 \rm shift} $ are the surface density contrast of the displace and shift catalog, respectively.  The default reconstruction method is by solving the full 2D transverse potential equation.

\begin{figure}
\centering
\includegraphics[width=0.98\linewidth]{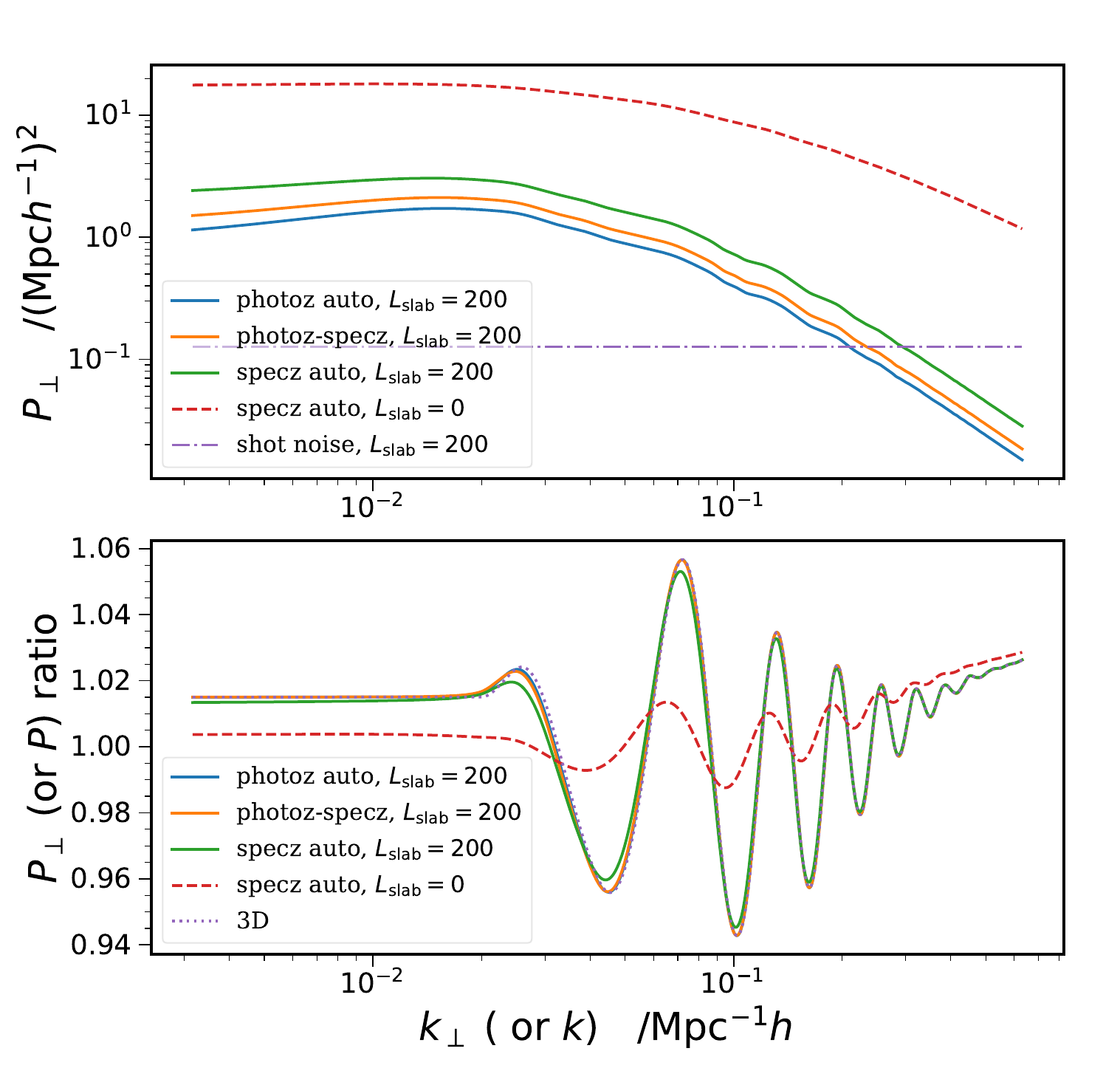}
\caption{ The linear matter transverse power spectrum for different types of data.
  In the upper panel, the photo-$z$ auto power spectrum (blue), the cross power spectrum between the photo-$z$  and  spec-$z$ data (orange), and the auto spec-$z$ power spectrum (green solid for $L_{\rm slab  } =200 \MpcOh$ and red dashed for $L_{\rm slab  } = 0 \MpcOh$) are compared. \change{ We also overplot the shot noise contribution for $L_{\rm  slab} = 200 \MpcOh $ (dotted-dashed, purple) expected from a typical photometric survey sample. }   In the lower panel, the ratios between the wiggle and the no-wiggle power spectrum for these data types are shown. \change{ In addition, the ratio from the linear 3D power spectrum (dashed) is also shown, and it coincides with spec-$z$  for $L_{\rm slab  } = 0 \MpcOh$ well. }  }
\label{fig:Pk_perp_auto_cross_specz}
\end{figure}

\begin{figure}
\centering
\includegraphics[width=0.98\linewidth]{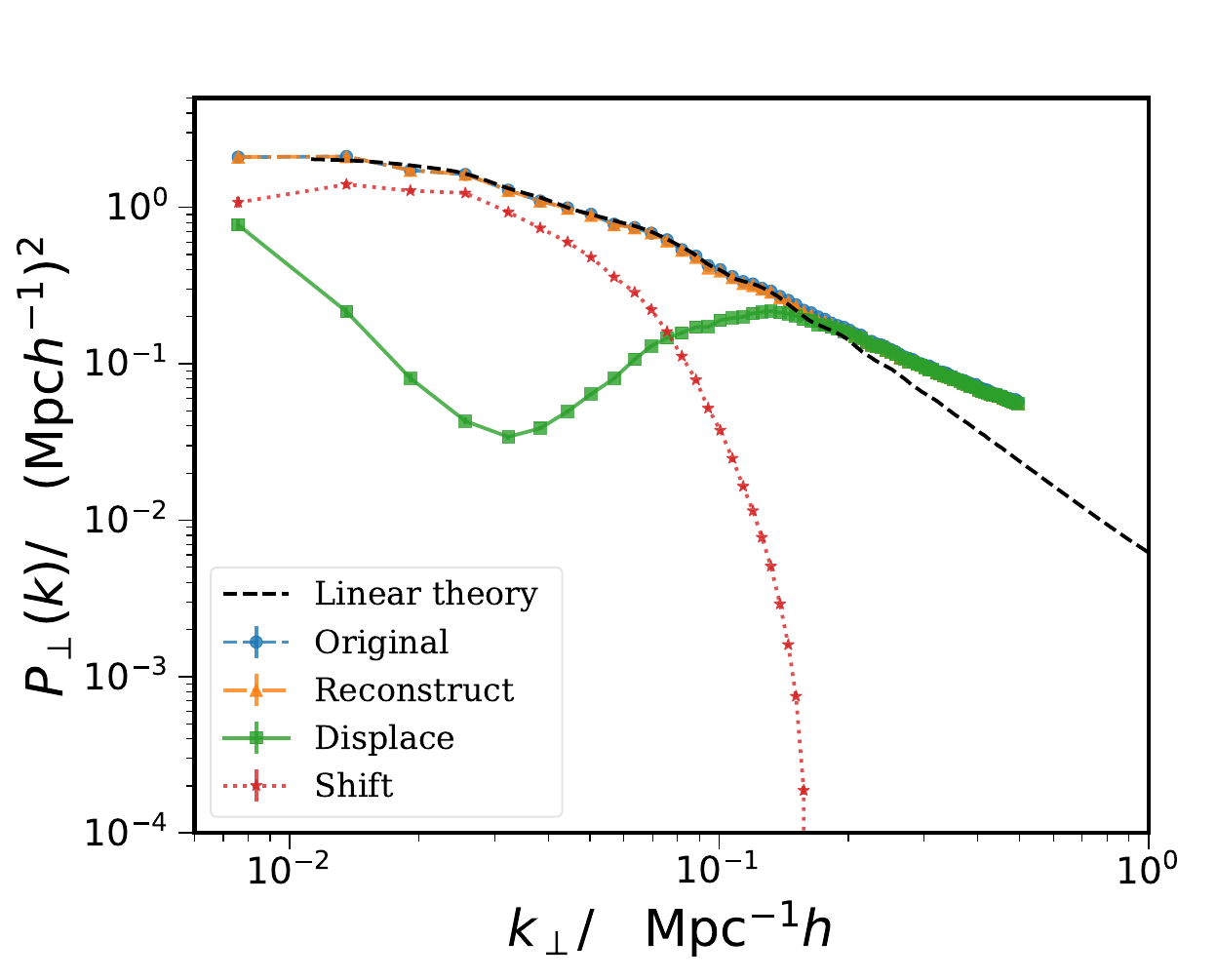}
\caption{ The matter transverse power spectrum at $z=0$. The original power spectrum (blue circles, almost completely covered by the reconstruction results), reconstructed power spectrum (orange triangles), displace catalog power spectrum (green squares), and shift catalog power spectrum (red stars) are numerical measurements, while the black dashed line shows the results from linear theory.   }
\label{fig:Pk_perp_DM_z0}
\end{figure}

\begin{figure*}
\centering
\includegraphics[width=0.98\linewidth]{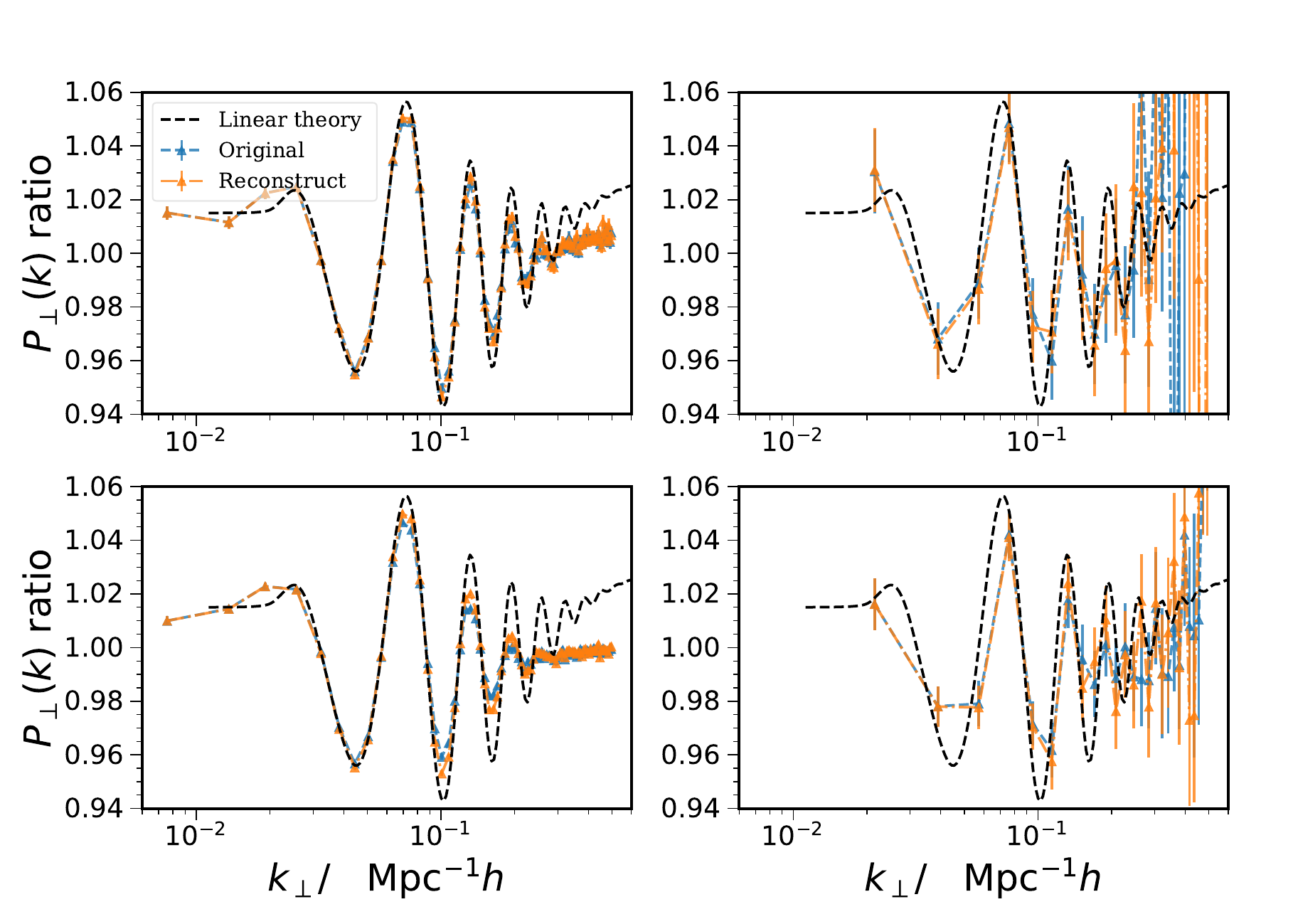}
\caption{ The ratio between the wiggle and the no-wiggle transverse power spectrum before (blue circles) and after (orange triangles) reconstruction.  The matter field results are shown on the left panels, while the right panels are for the halo field.  The results at $z=1$ (top) and 0 (bottom) are displayed. The ratio obtained with linear theory (black dashed) is overplotted for reference.  The matter field shows higher signal-to-noise and  there can be  half percent improvement in the BAO signals at $z=0$ after reconstruction.   The number density of the halo sample is too low to demonstrate the impact of BAO reconstruction.   } 
\label{fig:Pk_Wiggleratio_DMHalo_z1_0}
\end{figure*}

\subsubsection{Transverse power spectrum } 

To quantify the merits of the BAO reconstruction, we use the transverse power spectrum $P_\perp$, which is defined as
\beq
P_\perp ( k_\perp ) \Ddel( \bm{k}_\perp + \bm{k}_\perp' ) =  \langle \delta_2( \bm{k}_\perp )  \delta_2( \bm{k}_\perp' )  \rangle, 
\eeq
where $ \Ddel $ is the 2D Dirac delta function and $ \delta_2( \bm{k}_\perp )  $ is the Fourier mode of the 2D density contrast. Note that the angular brackets here denote ensemble averaging rather than tomographic bin averaging.   The transverse power spectrum can be taken to be  the flat sky limit of the angular power spectrum \citep{Dodelson_2003}, which is often used in the analysis of the photo-$z$ data.

In Appendix \ref{appendix:transverse_Pk_xi}, we show how $P_\perp$  is related to the underlying 3D power spectrum. In addition to deriving $P_\perp$ in the photo-$z$ case [Eq.~\eqref{eq:Pperp_p}], we also present the cross power spectrum between photo-$z$ data and spec-$z$ data in a slab [Eq.~\eqref{eq:Pperp_sp}], and the power spectrum for the spec-$z$ data in a slab [Eq.~\eqref{eq:Pperp_s}].  It is worth emphasizing that  the transverse power spectrum is generally suppressed by two factors: the photo-$z$ uncertainty and the averaging over the tomographic bin width.

To help understand the transverse power spectrum, we plot the linear theory results for different types of data in the upper panel of Fig.~\ref{fig:Pk_perp_auto_cross_specz}. The photo-$z$ uncertainty is modeled by a Gaussian window of width   $89.9 \MpcOh$, which corresponds to the photo-$z$ uncertainty we apply at $z=0$. The width of the slab is set to  $ L_{\rm slab} = 200 \MpcOh $.   The overall power of the photo-$z$ auto power spectrum is the lowest because it is suppressed by both the photo-$z$ uncertainty and the tomographic bin averaging. The cross power spectrum is the second lowest as it has one less power of photo-$z$ window suppression. The spec-$z$ auto power spectrum is the highest among them because it does not suffer from photo-$z$ suppression at all.  To demonstrate the impact of slab averaging, we show the spec-$z$ auto power spectrum with $ L_{\rm slab} = 0 \MpcOh $.  \change{ In this scenario, we have the truely transverse power spectrum 
  \beq
  \label{eq:Pperp_specLslab0}
P_\perp ( k_\perp ) = 2 \int_0^\infty d k_{\paral}  \,  P( k_\perp, k_{\paral} ),  
\eeq
which is obtained by setting $\sigma$ and $L$ in Eq.~\eqref{eq:Pperp_p} to zero.  }   The power in this case is much higher.  In general, we find that relatively mild averaging in the radial direction,  either due to the slab width or photo-$z$, can cause a sharp drop in power. When the radial averaging effect further increases, the drop tends to become  slower. For our parameter choices, the averaging is relatively strong, and so the sensitivity to the changes in averaging parameters are weak.

\change{ Here we would like to discuss the impact of the shot noise. In 2D, its contribution to the power spectrum is given by
\beq
P_{\rm shot} = \frac{ 1 }{ (2 \pi)^2  } \frac{ 1 }{ \bar{n}_2 } ,   
\eeq
where $ \bar{n}_2 $ is the 2D number density. In our setup, it is related to the 3D number density $ \bar{n}_3 $  as $ \bar{n}_2 = L_{\rm slab}  \bar{n}_3  $.  This means that the spec-$z$ with $L_{\rm slab} = 0 $ limit is not attainable in practice  because the shot noise blows up. In Fig.~\ref{fig:Pk_perp_auto_cross_specz}, we show the noise contribution for a  3D number density  $ \bar{n}_3  =  10^{-3} (\MpcOh)^{-3} $ and  $L_{\rm slab}=200\MpcOh $. This value of $ \bar{n}_3 $ is close to the level of current photometric survey sample, e.g.~\citet{CarneroRosell_etal2022}. The shot noise is subdorminant to the signal up to $ k \lesssim 0.2 \hOMpc $.  }


In the lower panel of Fig.~\ref{fig:Pk_perp_auto_cross_specz}, we highlight the BAO feature by plotting the ratio between the linear wiggle power spectrum and the no-wiggle one. The no-wiggle power spectrum is obtained from the Eisenstein-Hu initial conditions \citep{EisensteinHu_1998}. Although the linear theory results do not include BAO damping, it helps isolate the effect of photo-$z$ on the BAO.  We note that the BAO feature is strongest in photo-$z$ auto and cross power spectrum, while the spec-$z$ auto power spectrum shows weaker BAO  feature.   In fact, the spec-$z$ auto power spectrum with $ L_{\rm slab } = 0 \MpcOh $ even gives a shift in the BAO feature. The reason is that for the transverse power spectrum to be a good measure of the BAO feature, the BAO signals in the radial modes must by effectively erased. Otherwise, the radial modes will smear out the BAO feature and cause a shift in the BAO position. \footnote{ \change{  We also note that the ratio for the spec-$z$ result with $ L_{\rm slab } = 0 \MpcOh $ is lower than the other cases by about $1\% $ on large scale ($k \lesssim 0.03 \hOMpc $). This is because the amplitude of the no-wiggle  $P(k)$ is slightly lower than the wiggle one on large scale, and the diference diminishes on smaller scales.    This causes $ \sim  1.5 \% $ difference in the ratio on large scales. For spec-$z$ result with $ L_{\rm slab } = 0 \MpcOh $ (Eq.~\eqref{eq:Pperp_specLslab0}), without the window function ($\mathrm{sinc}^2$) suppression, it receives contribution from a large $k$ range, and hence  the ratio  between the two cases is reduced. } }  For the photo-$z$ auto and cross power spectrum, only the transverse modes contribute and hence they show the highest signal.  \change{ We have also plotted the ratio for the linear 3D power spectrum, it coincides with the photo-$z$ auto or cross power spectrum with $ L_{\rm slab } = 200 \MpcOh $  well. This reflects that the transverse BAO signals is the same as the 3D isotropic ones. }  However, we stress that in the measurement, \change{ the total signal-to-noise (S/N) matters, which is affected by the number of modes available and the S/N per mode.  The transverse measurements lose the radial modes relative to the 3D ones. }  Moreover, although the transverse power spectrum shows the highest signal, its power is low and its S/N is reduced.  Of course, for the spectroscopic case, the right statistic to use is the isotropic power spectrum in real space and the multipole power spectrum in redshift space.    In passing, we mention that the transverse power spectrum for photo-$z$ data carries a lot of similarities to the radial power spectrum for the 21 cm intensity mapping \citep{Villaescusa-Navarro:2016kbz}, in which only the radial BAO can be effectively measured due to poor angular resolution.

We show the pre- and post-reconstruction transverse power spectrum  measured from simulation in Fig.~\ref{fig:Pk_perp_DM_z0}. The results shown are for the matter field at $z=0$.  We have also plotted the power spectrum of the displace and the shift catalog. Indeed the power of the shift catalog is primarily in the low $k_\perp$ regime and the displace contains more small scale power. The \change{overall shape of} the power spectrum of the reconstructed catalog agrees with the original one well.  For reference, we have shown the linear transverse power spectrum predicted by Eq.~\eqref{eq:transverse_Pk}, and we find good agreement with simulation measurements on large scales.

\begin{figure}
\centering
\includegraphics[width=0.98\linewidth]{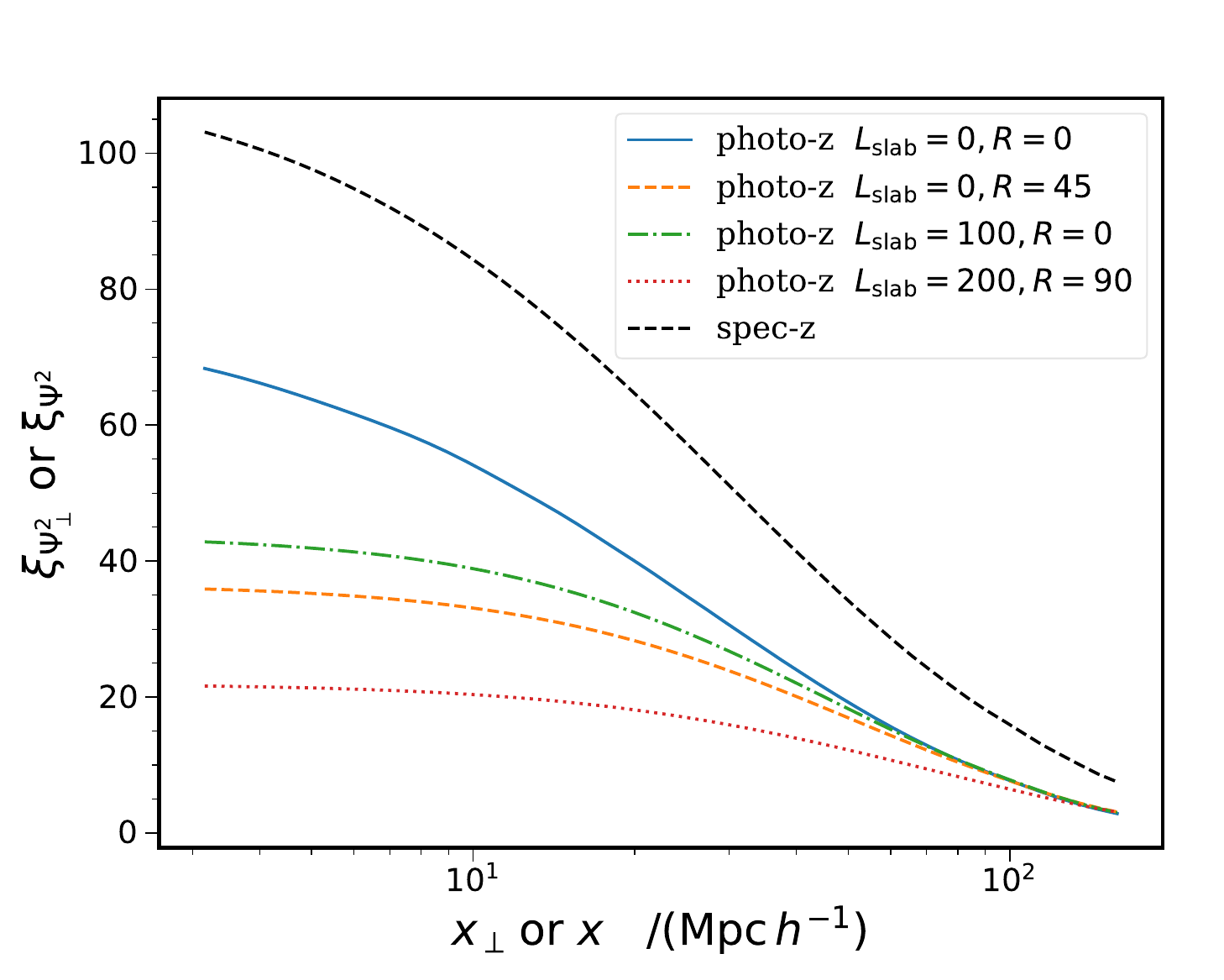}
\caption{ \change{ The correlation function of the transverse displacement field for different slab width $L_{\rm slab } $ and photo-$z$ smoothing length $R$: $L_{\rm slab }=0 \MpcOh $ and $R=0 \MpcOh $ (solid blue),  $L_{\rm slab }=0 \MpcOh $ and $R=45 \MpcOh $ (dashed orange),  $L_{\rm slab }=100 \MpcOh $ and $R=0 \MpcOh $ (dotted-dashed green), and   $L_{\rm slab }=200 \MpcOh $ and $R=90 \MpcOh $ (dotted red). The last one corresponds to our fiducial setup. The  correlation of the displacement field for the spec-$z$ case (dashed black) is shown for comparison, and it is higher than the photo-$z$ case even in the limit of  $L_{\rm slab }=0 \MpcOh $ and $R=0 \MpcOh $. }  } 
\label{fig:Psi_correlation}
\end{figure}

\subsubsection{ Performance of the reconstruction  }

We now move to check how much improvement that the BAO reconstruction brings. In Fig.~\ref{fig:Pk_Wiggleratio_DMHalo_z1_0}, we show the transverse power spectrum before and after BAO reconstruction. In order to highlight the BAO feature, we have divided by the no-wiggle power spectrum. We have shown the results for the matter and the halo, at $z=1$ and 0 respectively.  To guide the eyes, we have plotted the results obtained using linear theory. Without BAO damping in the linear theory, it sets the maximum signal that we can possibly recover.

The results for matter is shown on the left panels of Fig.~\ref{fig:Pk_Wiggleratio_DMHalo_z1_0} ($z=1$ on the top and $z=0$ at the bottom).  At $z=1$, the improvements are most obvious in the third and fourth peak. On larger scales the effect of nonlinear damping is milder and this limits the gain from BAO reconstruction, while on smaller scales,  reconstruction is not accurate enough to produce significant effect.  Reconstruction yields larger impact at $ z=0 $ because of stronger nonlinearity at low $z$. The effects are apparent in the second, third, and fourth peak. At maximum, the signals can be strengthened by half percent level.

We have also shown the corresponding results for the halo sample on the right panels of Fig.~\ref{fig:Pk_Wiggleratio_DMHalo_z1_0}.  We have maximized the number density of the halo sample by using all the halos with number of particles larger than 20.  
Moreover, compared to the matter case, we have increased the bin width in power spectrum measurement from $ 2 k_{\rm F} $  to  $ 6 k_{\rm F} $ ($  k_{\rm F} $ is the fundamental mode of the box) to reduce the impact of noise.  However, the signal-to-noise for the halo sample is still too low to reveal the impact of BAO reconstruction.

While direct comparison with the spec-$z$ reconstruction results are not possible, it seems that the improvment in the transverse BAO strength is less impressive than the spec-$z$ case. This is likely due to the coherence length of the displacement being lower than that in the spec-$z$ case. To look into this, we compare the correlation function of the displacement field for  the spec-$z$ and photo-$z$ case. For simplicity, we consider the linear theory results [see \cite{Chan2014} for the 1-loop case], and limit ourselves to the real space. For the spec-$z$ case, the correlation is given by
\begin{align}
  \xi_{\Psi^2}( x ) & = \langle \bm{\Psi}( \bm{r} ) \cdot  \bm{\Psi}( \bm{r}+ \bm{x} ) \rangle \nn \\
  & = \frac{ 4 \pi  }{ x } \int \frac{dk}{k} \sin(k x)   P(k) .
\end{align}
The correlation for the transverse displacement field in the presence of photo-$z$ error can be derived in a manner similar to the derivation of the transverse correlation function shown in Appendix \ref{appendix:transverse_Pk_xi}.  Analogously, we have
\begin{align}
  \xi_{\Psi_\perp^2}( x_\perp ) & = \langle \bm{\Psi}_\perp ( \bm{r}_\perp ) \cdot  \bm{\Psi}_\perp ( \bm{r}_\perp + \bm{x}_\perp ) \rangle \nn \\
& =   2 \pi \int_0^\infty d k_\perp \, k_\perp  P_{ \Psi_\perp^2 } (k_\perp ) J_0( k_\perp \Delta x_\perp ) , 
\end{align}
where $ P_{ \Psi_\perp^2 } $ reads
\begin{align}
    \label{eq:transverse_PkPsi2}
     P_{\Psi_\perp^2} (k_\perp )  & =   \int d k_{\paral}   ( 2 \pi )^2   \left|W \left(k_{\paral} \right)\right|^2     \mathrm{sinc}^2 \Big( \frac{  k_{\paral} L }{2} \Big) \frac{ k_\perp^2 }{ k^4 }   P( k_\perp, k_{\paral} ).
\end{align}
\change{ In Fig.~\ref{fig:Psi_correlation}, we compare $\xi_{\Psi_\perp^2} $ against $\xi_{\Psi^2} $. For $\xi_{\Psi_\perp^2} $, we have presented the results for different slab length, $ L_{\rm slab }$ and Gaussian smoothing scales due to photo-$z$, $R$.  Even in the limit of $ L_{\rm slab}= 0 \MpcOh $ and $ R = 0 \MpcOh $,  $\xi_{\Psi_\perp^2} $ is lower than $\xi_{\Psi^2} $.  In this case, if we further consider the limit $ x \rightarrow 0 $, it is easy to show that  $\xi_{\Psi_\perp^2} / \xi_{\Psi^2}  = 2 / 3 $. The ratio decreases as $x$ increases and reaches about 0.5 at $ x= 100 \MpcOh $.  If we crank up $ L_{\rm slab} $ and/or  $ R $, the ratio is even smaller.  $ L_{\rm slab}= 200 \MpcOh $ and $ R = 90 \MpcOh $ correspond to our fiducial setup. }  Thus we find that the photo-$z$ results are indeed much lower than the spec-$z$ case, supporting the argument that the weaker correlation of the displacement field is the cause of the less effective reconstruction results.

\subsection{ Reconstruction under different conditions}

We turn to test how  reconstruction works under  different conditions.   We focus on the matter field at $z= 0$ exclusively in the rest of the analysis for its strong signal-to-noise.


\subsubsection{ $\langle \phi_{\rm p} \rangle $  solutions } 
\label{sec:various_solution_check}

In this subsection, we test different methods to estimate the transverse displacement potential $ \langle \phi_{\rm p} \rangle  $.   First, in Sec.~\ref{sec:Direct3D_inversion}, we discuss that we can directly invert the 3D Poisson equation to obtain  $ \phi_{\rm p}( \bm{x}_\perp, x_{\paral}^{\rm p} ) $, and then project it to the transverse direction to yield  $ \langle \phi_{\rm p} \rangle ( \bm{x}_\perp ) $. Here we test if this method can effectively reconstruct the transverse BAO.   Second, we find in Sec.~\ref{sec:sources_magnitude} that the sources to the transverse potential equation is dominated by the surface density term. It is natural to consider the approximation scheme in which the surface density is the sole source term.

\begin{figure}
\centering
\includegraphics[width=0.98\linewidth]{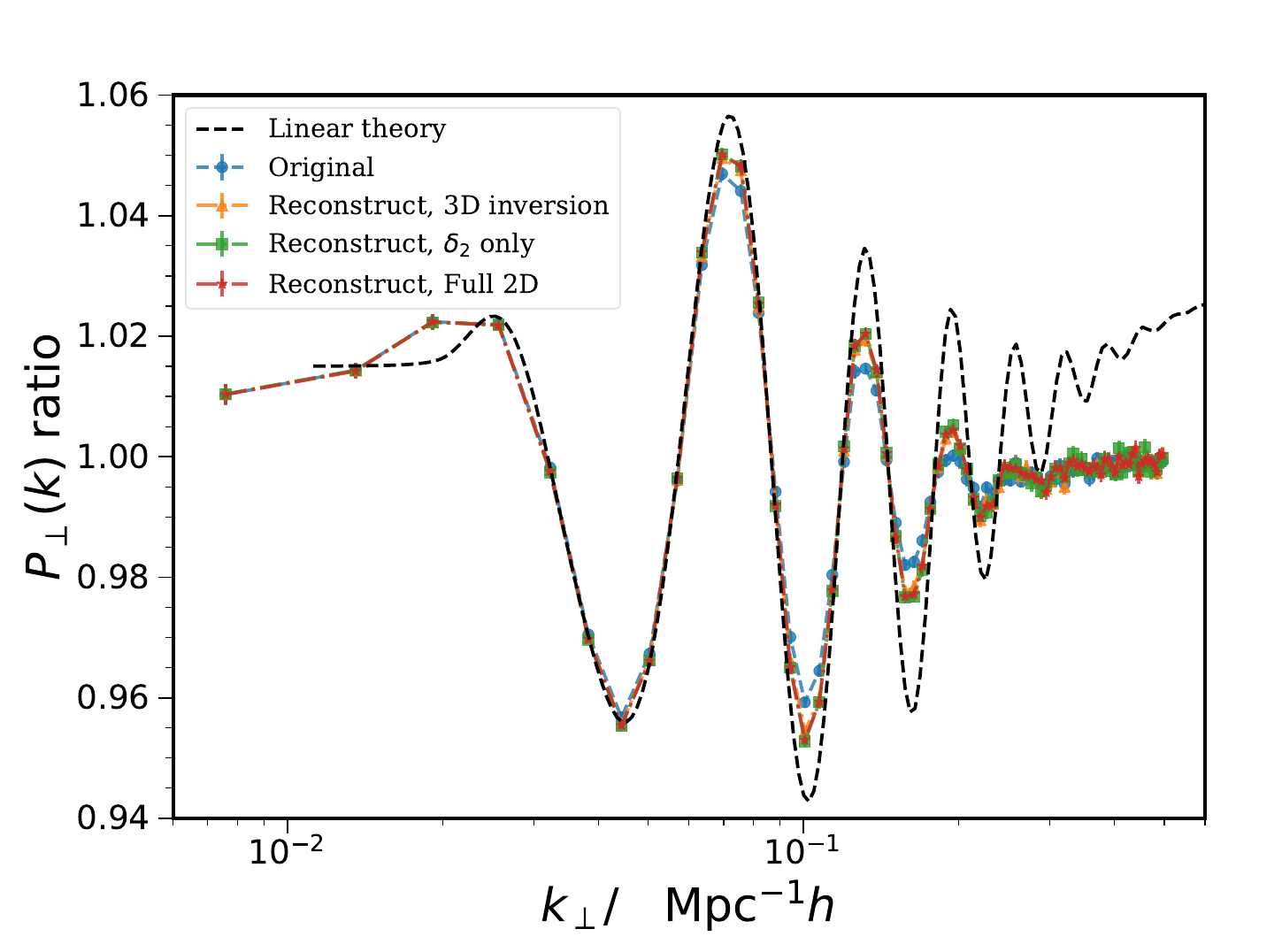}
\caption{ The ratio between the wiggle and no-wiggle transverse power spectrum before (blue circles) and after  reconstruction by different methods: inversion of the 3D Poisson equation (orange triangles), using the surface density as the sole source term (green squares), and the full 2D transverse potential equation (red stars).  Here and for the rest of the performance test plots, the results are based on the matter field at $z=0$.  } 
\label{fig:Pk_Wiggleratio_DM_z0_methodcompare}
\end{figure}

In Fig.~\ref{fig:Pk_Wiggleratio_DM_z0_methodcompare}, we compare the results obtained from these two methods against the full 2D equation result, which has been the fiducial choice in our analysis.  Different methods give very similar results. The observation that  the density-only case agrees with the full 2D solution well demonstrates that the radial potential term is indeed negligible compared to the surface density. Although the motion is inherently 3D, the transverse reconstruction is completely dominated by the density on the slab.  This approximation further simplifies the reconstruction procedures and makes them resemble the spectroscopic ones even more.

The fact that the reconstruction is dominated by the surface term also help explain why the transverse potential equation results agree with the 3D Poisson equation ones  well.  The key difference between these methods is  the order of projection and reconstruction solution. It is useful to contrast this case with the difference between the angular correlation function and the projected 3D correlation function discussed in \cite{Chan_xip2022}.  In the  angular correlation function  measurement, one first projects the data to the angular space and then performs the correlation function measurement.  For the projected 3D correlation function, we regard the photo-$z$ data as a 3D distribution to first perform the 3D correlation measurement, and then project it to the transverse direction to suppress the anisotropies induced by photo-$z$. It was found that both statistics generally yield similar results, but projected 3D correlation function sometimes can be more sensitive to the photo-$z$ noise than the angular one. The reason is that in the angular correlation function treatment, by first projecting the data to the angular space, it is more effective in nulling the photo-$z$ contamination.   The smallness of the radial contribution implies that the impact of the photo-$z$ uncertainties would be further diminished at the field level, in contrast to the correlation function.  



\subsubsection{ Photo-$z$ uncertainties } 

\begin{figure}
\centering
\includegraphics[width=0.98\linewidth]{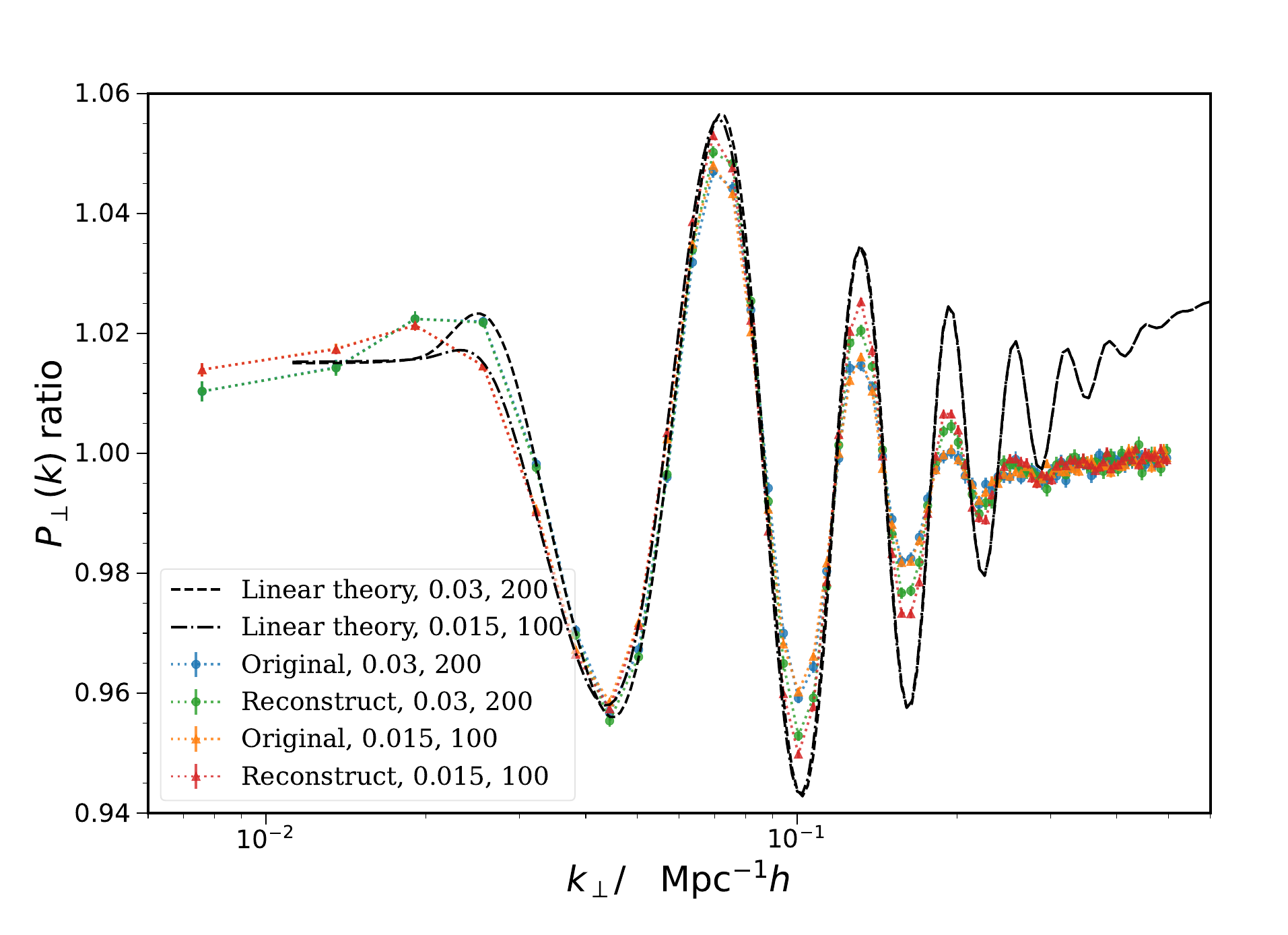}
\caption{ The ratio between the wiggle and the no-wiggle transverse power spectrum before and after reconstruction for $\sigma_z=0.03$ (slab width of 200  $\MpcOh$, circles) and  $\sigma_z=0.015$ (slab width of 100  $\MpcOh$, triangles). The linear theory results for these two cases are shown as dashed and dotted-dashed curves, respectively.  While there are only small differences for the pre-reconstruction case (blue for 0.03 and orange for 0.015) in the range of  $k_\perp > 0.05 \hOMpc $, the reconstructed  field for $\sigma_z =0.015 $ (orange) show more pronounced enhancement than  $\sigma_z =0.03 $ (green) thanks to smaller  $\sigma_z$ and thinner slab width. }
\label{fig:Pk_Wiggleratio_DM_z0_sigmaz_Dz}
\end{figure}

Here we check how the reconstruction fairs when the level of photo-$z$ uncertainty varies. In Fig.~\ref{fig:Pk_Wiggleratio_DM_z0_sigmaz_Dz}, we compare the reconstruction results obtained with  $\sigma_z=0.015$  against those with  $\sigma_z=0.03$.  The corresponding comoving smoothing scales are $ 45.0 \MpcOh $ ($\sigma_z=0.015$) and  89.9 $\MpcOh$ ($\sigma_z=0.03$), respectively.  As the slab width is chosen to be similar to the scale of the photo-$z$ uncertainty,   with a smaller $\sigma_z$, we can afford to use a thinner slab.  For $\sigma_z=0.015$, we utilize a slab width of $100 \MpcOh$ instead of the fiducial value of $200 \MpcOh$. This allows us to reduce the damping in power due to the averaging over the slab width.

Because of different comoving photo-$z$ smoothing scales and slab width, there are small differences between the two cases on large scales ($k \lesssim 0.05 \hOMpc $). These are also noticeable in the linear theory results.   On smaller scales, the linear theory results coincide with each other.  Numerically,  we find that the ratios for the pre-reconstruction case are indeed very similar,  with the  BAO signals from the $\sigma_z=0.015$ case stronger by a very small margin only.   These show that $ \sigma_z=0.015 $ is large enough to eliminate the radial modes so that there is little difference from the $ \sigma_z=0.03 $ case.

The impact of reconstruction on these two cases are also very similar, but we find that the  $\sigma_z=0.015$ case results in  stronger BAO signal after reconstruction.  Both the reduction in  $\sigma_z $ and the simultaneous shrinkage of the slab width contribute to the enhancement. This shows that for weaker photo-$z$ contamination, the effect of the BAO reconstruction is more pronounced.


\subsubsection{ Smoothing scales  }

The fiducial setup utilizes a Gaussian window with width of 15 $\MpcOh$.  This resembles the smoothing scale commonly utilized in spectroscopic BAO reconstruction, where it is often chosen within the range [10, 20] $\MpcOh$  \citep{Burden_etal2014}. In this subsection, we test the impact of different smoothing scales.

\begin{figure}
\centering
\includegraphics[width=0.98\linewidth]{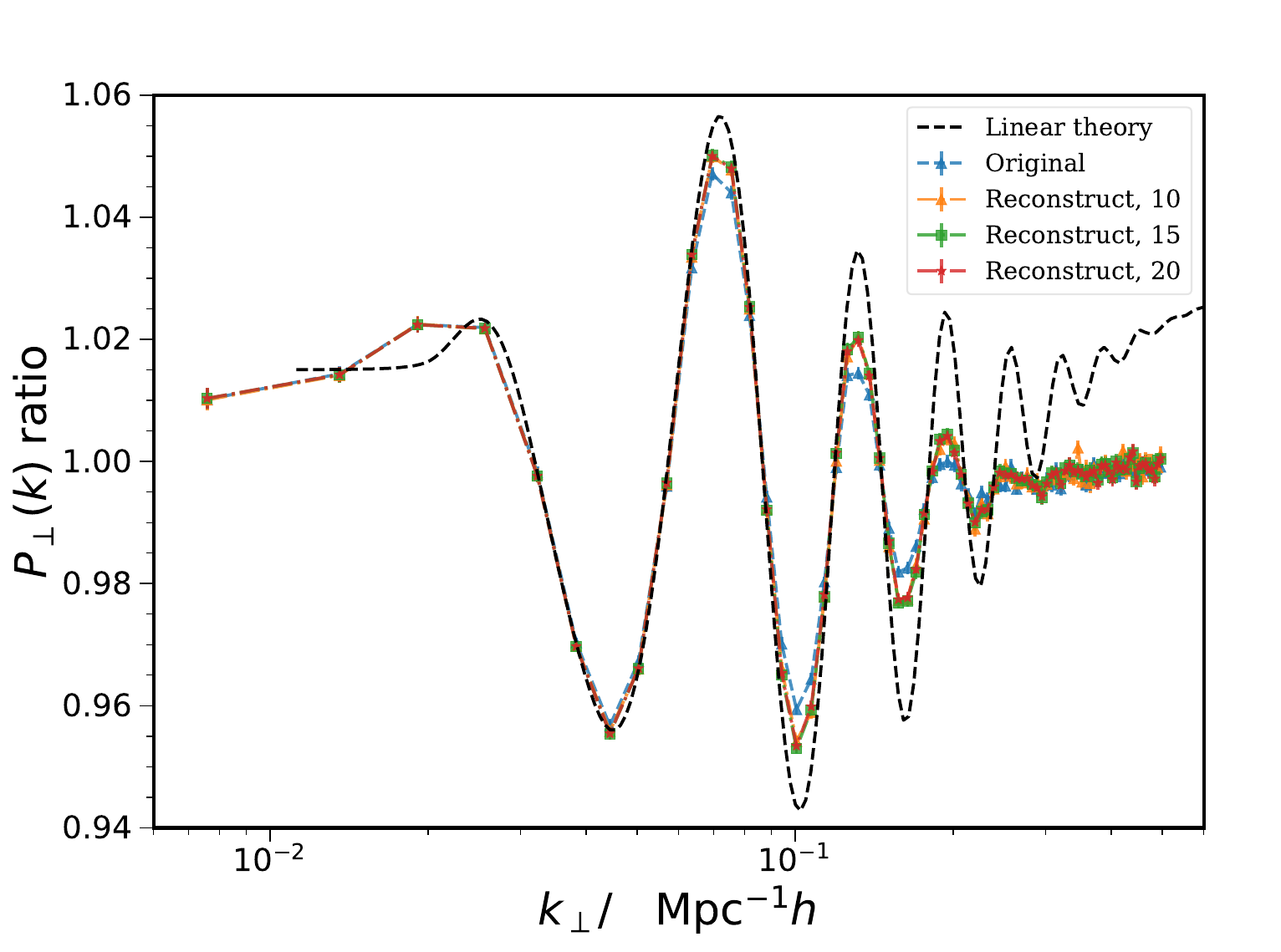}
\caption{ The ratio between the wiggle and no-wiggle transverse power spectrum before (blue circles) and after reconstruction.  We compare the performance of the reconstruction algorithm obtained with three different Gaussian smoothing scales: 10 (orange triangles), 15 (green squares), and 20 $\MpcOh $ (red stars).     } 
\label{fig:Pk_Wiggleratio_DM_z0_smoothingRG}
\end{figure}

In Fig.~\ref{fig:Pk_Wiggleratio_DM_z0_smoothingRG}, we compare the reconstruction results obtained with three different smoothing scales: 10, 15, and 20 $\MpcOh$.  Overall, the reconstruction results obtained with different smoothing scales are similar.  In fine details, for the fourth and higher peaks, small smoothing scale ($ 10 \MpcOh$) results in slightly noisier results, and the larger smoothing scale ($ 20 \MpcOh$) leads to  mildly stronger power suppression.  Although the results are not overwhelming, these seem to suggest that  $15 \MpcOh$  is able to suppress the small-scale noise without excessive power reduction.

One may worry that the tests here are conducted on the dark matter field and a larger smoothing scale may be required to get the optimal result because the influence of the stochasticity is minimized in the case of matter. We try to test this by downsampling the matter density field, e.g.~by a factor of 10, to a number density of $ 0.013 \,  (\MpcOh)^{-3} $. However, even in this case, the noise introduced is so large that the putative differences among them cannot be uncovered.

\subsubsection{ Real space versus redshift space}

So far we have been showing the results in redshift space. For completeness, we compare the reconstruction results in real space with that in redshift space.  The linear RSD \citep{Kaiser87} mainly enhances the power of the transverse power spectrum at relatively low $k_\perp$,  $k_\perp \lesssim 0.04 \hOMpc $.  At higher $k_\perp$, the impact could be modeled phenomenologically in the dispersion model by a Gaussian or Lorenzian damping factor \citep{PeacockDodds_1994}. This damping effect is degenerate with the much larger photo-$z$ damping, and thus it can be effectively absorbed into the photo-$z$ damping term.

\begin{figure}
\centering
\includegraphics[width=0.98\linewidth]{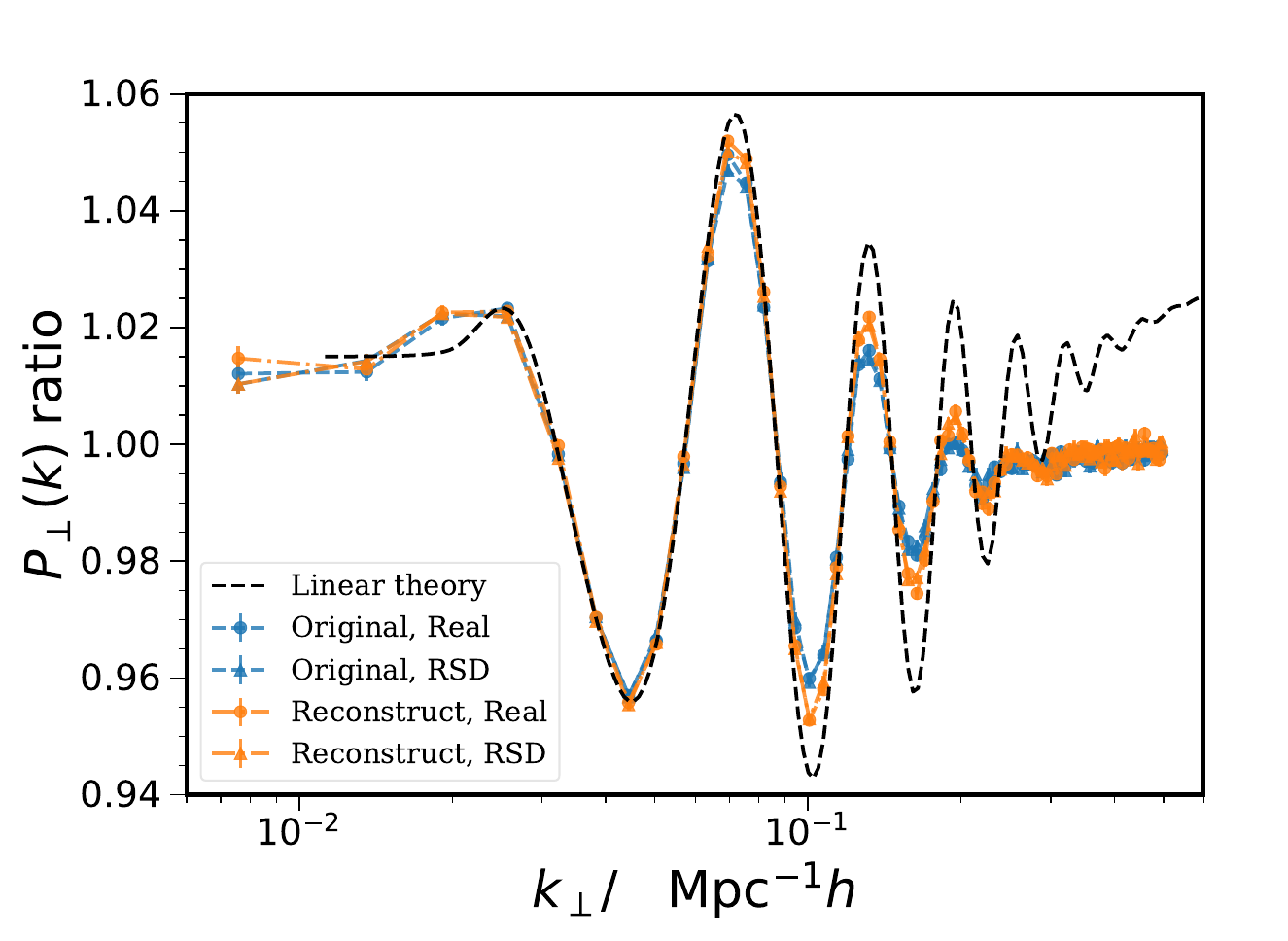}
\caption{ The ratio between the wiggle and no-wiggle transverse power spectrum before (blue) and after (orange) reconstruction in real space (circles) and in redshift space (triangles).  } 
\label{fig:Pk_Wiggleratio_DM_z0_Real_RSD}
\end{figure}

We compare the results in real space and in redshift space in Fig.~\ref{fig:Pk_Wiggleratio_DM_z0_Real_RSD}.  The results are very similar in these two spaces, irrespective of it is before or after reconstruction.   This is  expected given we perform transverse reconstruction. 

\subsection{ Cross correlation with the initial conditions }
\label{sec:Propagator}

For the spectroscopic data, the cross correlation between the Eulerian field and  the linear field, which is also called the propagator, is often used to quantify the information gain from reconstruction. Explicitly, it is defined as   \citep{CrocceScoccimarro2006,CrocceScoccimarro_2008}
\begin{align}
G(k,z) = \frac{ \langle  \delta_{\rm L}( k, z) \delta_X (-k,z) \rangle }{  \langle  \delta_{\rm L}( k, z) \delta_{\rm L} (-k,z) \rangle },
\end{align}
where $ \delta_{\rm L} $ is the linear density field and $ \delta_X  $ is the Eulerian field that we are interested in.   Inspired by this definition, we apply $G$ to test the merits of transverse reconstruction. For matter field in linear theory, the numerator is given by Eq.~\eqref{eq:Pperp_sp} and  the denominator  by  Eq.~\eqref{eq:Pperp_s}. Because  the numerator  $ P_\perp^{\rm ps} $ has one more power of the Gaussian window due to photo-$z$ contamination, the propagator is still lower than  unity even in the low $k_\perp$ limit.  \change{ Alternatively we can consider the cross correlation coefficient $r$ [e.g.~\citet{LiuYuLi_2021}]. However, due to finite range of integration over $ k_{\paral} $, the photo-$z$ and slab averaging effects do not cancel out exactly, $r$ also does not tend to unity in the low $k_\perp$ limit.  }

\begin{figure}
\centering
\includegraphics[width=0.98\linewidth]{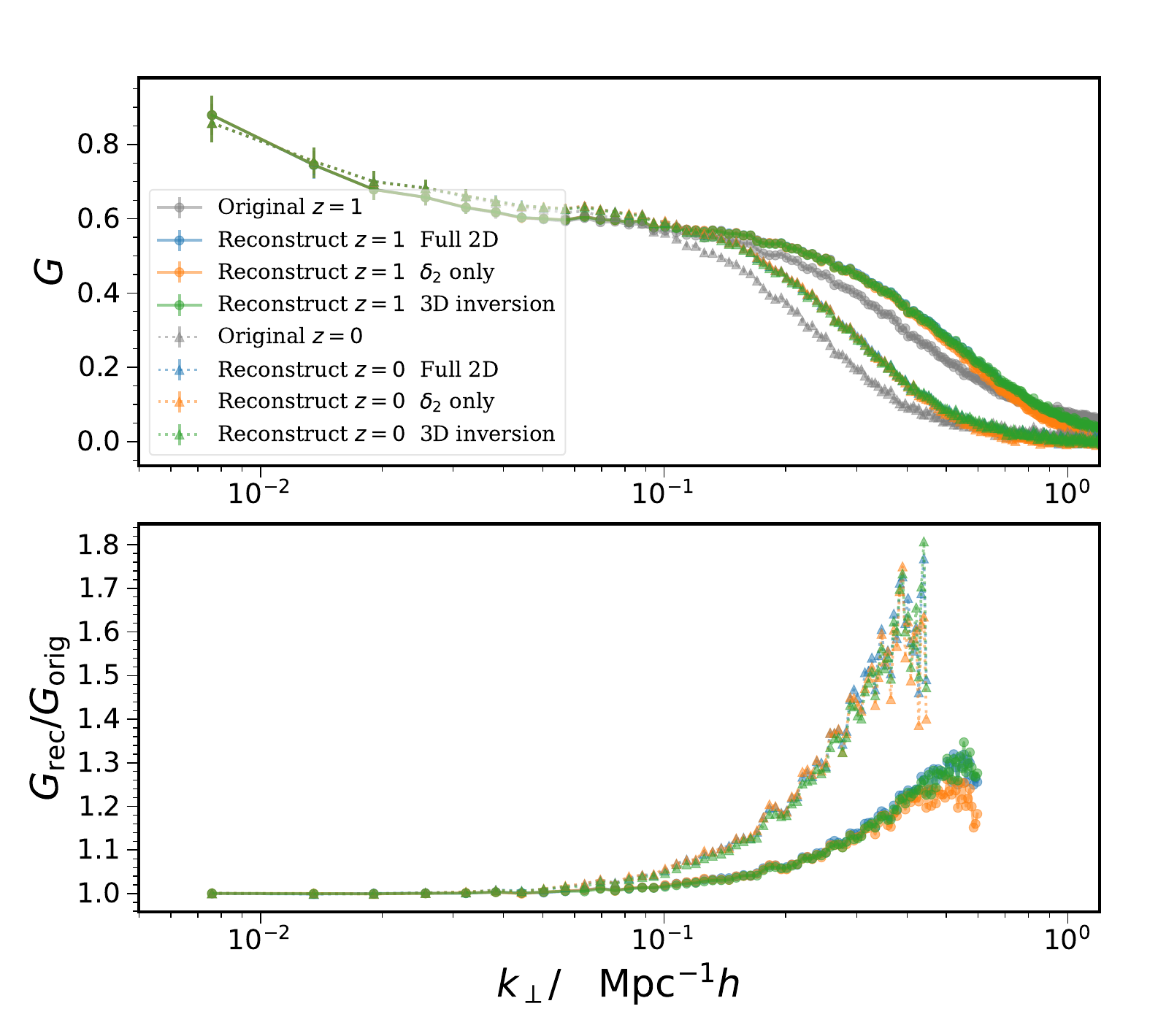}
\caption{ The upper panel shows the  propagators for the pre- (grey) and post-reconstruction field, while the lower one displays the ratios of  the reconstructed field propagator to the original field one.  Both the results at $z=1$ (circles with solid lines) and 0 (triangles with dotted lines) are shown.  The results from three reconstruction methods are compared: full 2D (blue) and density-only (orange) transverse potential equation, and the direct 3D Poisson equation inversion (green).  In the ratio plot, we stop showing the data points when the scatter becomes large.   These reconstruction methods yield overall similar enhancements in correlation in the intermediate range of scales  $ 0.1 \hOMpc  \lesssim k_\perp  \lesssim 0.6 \hOMpc $. }
\label{fig:PropagatorG_DM_z0_1}
\end{figure}

In Fig.~\ref{fig:PropagatorG_DM_z0_1}, we compare the propagator for the original field and the reconstructed field. We have shown the results for the matter field at $z=1$ and 0. To highlight the amount of increase in correlation, we have shown the ratio between $G$ from the reconstructed field and that from the original field in the lower panel as well. For clarity, we stop showing the data when the scatter becomes large.    For $k_\perp \lesssim 0.1 \hOMpc$, the propagator does not show any changes before and after reconstruction, consistent with the ratio test conducted in the previous subsections.  At low  $k$, the $z=0$ results are higher than the $z=1$ case because the photo-$z$ uncertainty scale applied at $ z=1 $ is larger.   In the intermediate scales, $ 0.1 \hOMpc  \lesssim k_\perp  \lesssim 0.6 \hOMpc $,  $G$ from the reconstructed field is higher than the original one, indicating that the reconstruction algorithm indeed moves the particles closer to the initial position.   For $ z=0 $, the correlation is increased by 50\% at  $ k_\perp = 0.3 \hOMpc $, and by 25\%  at  $ k_\perp = 0.4 \hOMpc $ for $z=1$.   In this range of scales,  the cross correlation  at $z=1 $  is stronger than at $z=0 $.   We can explain this if the increase in the correlation at $ z=1 $  is large enough to compensate the stronger photo-$z$ damping that it suffers. However, in terms of the fractional change in amplitude, the gain from reconstruction is more appreciable at $z=0  $ than $z=1$ because of stronger nonlinearity at $z=0$, in agreement with reconstruction test in previous subsections. On smaller scales,  $ k_\perp \gtrsim 0.6 \hOMpc $,  $G$ from the reconstructed field is even lower than the original one.  On such small scales, the reconstruction displacements are not correct and hence they essentially give noise.  Overall, the three different methods yield very similar results, consistent with the ratio test results.   


In contrast, for the spec-$z$ data, the increase in $G$ is much more appreciable. For example, at $z=1$ and  $k= 0.4 \hOMpc $, Fig.~7 of \cite{Seo_etal2010} shows that in redshift space $G$ is increased by a factor 3 after reconstruction. While the benefits of reconstruction are less attractive for photo-$z$ data, there is sizable increase in the correlation compared to no reconstruction at all.

We mentioned that  Eq.~\eqref{eq:phipp_smoothintg} incorrectly estimates the radial contribution term. Although this is not obvious for the performance tests conducted in Sec.~\ref{sec:ReconstructionTests}, it becomes clear for $G$. If we use Eq.~\eqref{eq:phipp_smoothintg} for the radial term instead, we would find that the full 2D solution yields an estimate of $G$ lower than the other two methods at $z=0$, and even lower than the pre-reconstruction case at $z=1$.  This in turn suggests that the propagator is more telling on the information gain than the power spectrum ratio.


\section{Conclusions}
\label{sec:conclusions}

Recently, we have seen a flurry of robust transverse BAO measurements using photometric data.  With the ongoing and upcoming large-scale structure photometric surveys mentioned in the Introduction, it is anticipated that photometric BAO measurements will play a more prominent role.  While the reconstruction technique has been extensively employed to enhance the significance of BAO detection in spectroscopic data analysis, its application to photometric data is still lacking.  This work aims to fill this gap.

We lay down the formalism in Sec.~\ref{sec:formalism}.  We first present a direct generalization of the conventional Zel'dovich reconstruction by solving a 3D Poisson equation using the position in photo-$z$ space.  The transverse potential follows from the projection of the 3D potential to the transverse direction.  A second approach consists in isolating the transverse and the radial components to formulate a transverse potential equation, with the surface density and the radial potential serving as the source terms.  This approach makes clear that the surface density term is completely dominant over the radial potential term.   In Sec.~\ref{sec:NumericalAnalysis}, we proceed to evaluate the performance of the reconstruction algorithm through comoving $N$-body simulations.  For the matter field, we find that the transverse BAO reconstruction can indeed raise the BAO signals, resulting in a notable half-percent increase in BAO strength at $z=0$.  Besides, BAO reconstruction enhances the cross correlation with the initial field, with a 50\% increase observed at $k_\perp = 0.3 \, \hOMpc$ at $z=0$.

Subsequently, we examine the performance of the reconstruction under various conditions.
We compare the displacement potentials obtained through three different methods: direct 3D Poisson equation inversion, the complete 2D transverse potential equation solution, and the density-only 2D solution. Remarkably, these approaches yield similar BAO reconstruction results, indicating that we can streamline the procedures by treating the problem as a 2D one.   Moreover, if the level of photo-$z$ uncertainty is reduced, the benefits of reconstruction are  more pronounced.  With the reduction of the photo-$z$ uncertainty, we can use a narrower slab width and hence the transverse displacements are made more accurate.  
Analogous to the spec-$z$ case, we observe that a smoothing scale of $15 \MpcOh$ effectively mitigates small-scale noise without a significant reduction in power. In transverse reconstruction, the outcomes remain insensitive to whether it is carried out in real space or redshift space.


After demonstrating the effectiveness of the transverse reconstruction, our future plans include applying this method to current photometric survey datasets, such as the DES Y6.
The method's simplicity and its resemblance to spec-$z$ analysis facilitate its straightforward adaptation. Nonetheless, before proceeding, we intend to conduct further validation of the method using more realistic mock catalogs. The DES Y3 \citep{DES:2021fie} or Y6 mock catalogs exhibit a number density that surpasses the halo sample attempted in this study by an order of magnitude. Therefore, the challenges associated with the lower number problem encountered in this work are anticipated to be alleviated for the DES mocks.  While we work in the plane parallel limit here, for large survey dataset, we have to generalize the formalism to incorporate the curved sky effect. We will present the checks elsewhere.    Moreover, we expect that the advanced methods mentioned in the Introduction can be readily adapted for transverse reconstruction.

\section*{Acknowledgments}
We thank Yanchuan Cai, Zhejie Ding, Yu Yu, Zhongxu Zhai, and Pengjie Zhang for useful discussions. \change{ We are also grateful to the anonymous referee for his/her constructive comments that improve the presentation of the paper. } This work is supported by  the National Science Foundation of China under the grant number 12273121 and the science research grants from the China Manned Space Project.

\section*{Data availability}
The data underlying this article will be shared on reasonable request to the corresponding author.

\bibliographystyle{mnras}
\bibliography{references}

\appendix

\section{ Transverse power spectrum and transverse correlation function  }
\label{appendix:transverse_Pk_xi}

Here we show how the transverse power spectrum and the transverse correlation function are related to the underlying three-dimensional power spectrum.  We start with the density contrast in photo-$z$ space:
\begin{align}
    &   \delta_{\rm p} \left(\bm{x}_{\perp}, x_{\paral}^{\rm p}  \right)  \nn \\
  = &  \int d x^{\prime}_{\paral} \, W (x_{\paral}^{\rm p} -x_{\paral}^{\prime} ) \delta (\bm{x}_\perp, x_{\paral}^{\prime} )  \nn \\
  = &  \int d x_{\paral}^{\prime} \int d p_{\paral} e^{i p_{\paral}\left(x_{\paral}^{\rm p} -x_{\paral}^{\prime}\right)} W\left(p_{\paral}\right)  \int d \bm{k}_\perp d k_{\paral}  \nn \\
    & \times  e^{i \bm{k}_\perp \cdot \bm{x}_\perp } e^{ i k_{\paral} x_{\paral}^{\prime}} \delta\left( \bm{k}_{ \perp }, k_{\paral}\right)  \nn \\
  = &  \int d \bm{k}_{\perp} e^{i \bm{k}_\perp \cdot \bm{x}_\perp } \int d k_{\paral} e^{i k_{\paral} x_{\paral }^{\rm p}    }   \delta\left( \bm{k}_\perp, k_{\paral}\right) [ 2 \pi W\left(k_{\paral}\right) ]. 
\end{align}
In practice we consider the 2D density in a tomographic bin. Its effect is to average  $\delta_{\rm p} $  over the tomographic bin as
\begin{align}
  \delta_{\rm p}^{(2)} \left(\bm{x}_{\perp} \right) &  = \frac{1}{L}  \int_0^L d x_{ \paral }^{\rm p}  \delta_{\rm p} \left(\bm{x}_{\perp}, x_{\paral}^{\rm p}  \right) ,
\end{align}
where  $L$ is the width of the tomographic bin.   After performing the integral over $ x_{ \paral }^{\rm p} $, we get
\begin{align}
  \delta_{\rm p}^{(2)} \left(\bm{x}_{\perp} \right) &  =  \int d \bm{k}_{\perp} e^{i \bm{k}_\perp \cdot \bm{x}_\perp } \int d k_{\paral}   e^{i \frac{  k_{\paral} L }{2}  } \mathrm{sinc}\, \frac{  k_{\paral} L }{2}   \nn \\
  &  \times \delta\left( \bm{k}_\perp, k_{\paral}\right) [ 2 \pi W\left(k_{\paral}\right) ]. 
 \end{align}

The transverse correlation function can be written as
\begin{align}
   &\left\langle \delta_{\rm p}^{(2)} \left(\bm{x}_\perp \right)    \delta_{\rm p}^{(2)} \left(\bm{x}_{\perp}^{\prime}\right)\right\rangle  \nn  \\
  = &     \int  d \bm{k}_\perp  e^{i \bm{k}_\perp \cdot \left(\bm{x}_\perp- \bm{x}_\perp^{\prime}\right)}    \int d k_{\paral}   ( 2 \pi )^2   \left|W \left(k_{\paral} \right)\right|^2 \nn \\
   & \times \mathrm{sinc}^2\,  \frac{  k_{\paral} L }{2}     P(k_\perp, k_{\paral} ), 
\end{align}
where $P $ is the redshift space power spectrum. For example, for halos, it is given by 
\beq
\label{Eq:RSD_Pk} 
 P(k_\perp, k_{\paral} )  = b_{\rm g}^2  \Big(  1 + f    \frac{  k_{\paral}^2  }{ k_\perp^2  + k_{\paral}^2  }    \Big)^2       P_{\rm lin} \Big(\sqrt{k_\perp^2+k_{\paral}^2}\Big) , 
\eeq
with $  P_{\rm lin} $ being the linear matter power spectrum.  It follows that  the transverse power spectrum reads
\begin{align}
    \label{eq:transverse_Pk}
  P_{\perp} (k_\perp )  & =   \int d k_{\paral}   ( 2 \pi )^2   \left|W \left(k_{\paral} \right)\right|^2     \mathrm{sinc}^2\, \frac{  k_{\paral} L }{2}     P( k_\perp, k_{\paral} ).
\end{align}
From Eq.~\eqref{Eq:RSD_Pk}, we see that the linear RSD increases  the power spectrum for small $k_\perp $. In our case, it only shows significant enhancement for  $k_\perp \lesssim 0.04 \hOMpc $.   On the other hand, there are two power suppression factors in Eq.~\eqref{eq:transverse_Pk}: the photo-$z$ window $ ( 2 \pi )^2   \left|W \left(k_{\paral} \right)\right|^2 $ and  $  \mathrm{sinc}^2 $ factor due to averaging over the tomographic bin. Both of them reduce power on all $k_\perp$ scales. 

Taking $W$ to be the Gaussian window $W_{\rm G}(k_{\paral} ) = \frac{1}{2 \pi} e^{-\frac{1}{2}  k_{\paral}^2 \sigma^2}$,
\beq
  \label{eq:Pperp_p}
P_{\perp} \left(k_\perp \right) = 2 \int_0^{\infty}  d k_{\paral}  e^{-k_{\paral}^2 \sigma^2 }   \mathrm{sinc}^2\, \frac{  k_{\paral} L }{2}    P  (k_\perp, k_{\paral} ).
\eeq
The correlation function is related to the power spectrum by a Hankel transform
\beq
\xi_\perp ( \Delta  x_\perp ) =  2 \pi \int_0^\infty d k_\perp \, k_\perp  P_\perp (k_\perp ) J_0( k_\perp \Delta x_\perp ) .  
\eeq

As a corollary, we can easily write down the transverse power spectrum for the spectroscopic data in a tomographic bin.   The cross power spectrum between photometric data and spectroscopic data in the bin is given by
\begin{align}
    \label{eq:Pperp_sp}
  P_{\perp}^{\rm ps} (k_\perp )  & =   \int d k_{\paral}   ( 2 \pi )  \left|W \left(k_{\paral} \right)\right|     \mathrm{sinc}^2\, \frac{  k_{\paral} L }{2}   P  (k_\perp, k_{\paral} ).
\end{align}
The auto power spectrum of the spectroscopic data in the bin reads
\begin{align}
  \label{eq:Pperp_s}
  P_{\perp }^{ \rm s} (k_\perp )  & =   \int d k_{\paral}  \,   \mathrm{sinc}^2\, \frac{  k_{\paral} L }{2}    P  (k_\perp, k_{\paral} ).
\end{align}

\bsp	
\label{lastpage}
\end{document}